\documentclass[11pt]{article}
\usepackage{jheppub}
\usepackage{tikz}

\usepackage{amsmath}
\usepackage{amssymb}
\usepackage{braket}
\usepackage{bbold}
\usepackage{pifont}
\usepackage{color}
\usepackage{physics}
\usepackage{wasysym}
\usepackage{bm}
\usepackage{slashed}
\usepackage{cases}

\usepackage[nottoc]{tocbibind}

\usepackage[centertableaux, smalltableaux]{ytableau}

\def\beq{\begin{equation}}
\def\eeq{\end{equation}}

\def\Tr{\mathop{\rm Tr}}

\fontfamily{cmss}

\def\crbig{\\\noalign{\vspace {3mm}}}

\newcommand{\md}{\mathrm{d}}
\def\ov{\overline}

\title{A stringy massive double copy}

\author[\eighthnote,\twonotes]{Dieter L\"ust}
\author[\sharp\quarternote]{, Chrysoula Markou}
\author[\twonotes]{, Pouria Mazloumi}
\author[\twonotes]{, Stephan Stieberger}

\affiliation[\eighthnote]{Arnold Sommerfeld Center for Theoretical Physics, \\ Ludwig--Maximilians--Universit\"at M\"unchen \\
Theresienstra{\ss}e 37, 80333 Munich, Germany} 

\affiliation[\twonotes]{Max--Planck--Institut f\"ur Physik (Werner--Heisenberg--Institut)\\
F\"ohringer Ring 6, 80805 Munich, Germany}

\affiliation[\sharp\quarternote]{Service de Physique de l'Univers, Champs et Gravitation, Universit\'e de Mons - UMONS \\
20 Place du Parc, B-7000 Mons, Belgium}

\emailAdd{dieter.luest@lmu.de, chrysoula.markou@umons.ac.be} \emailAdd{pmazlomi@mpp.mpg.de,stephan.stieberger@mpp.mpg.de}

\abstract{We derive a massive double copy construction within string theory. To this end, we use massive vectors of the open string spectrum that appear in compactifications to four dimensions and construct massive spin--2 tensors as closed string states, thereby mimicking the structure of the massless graviton. We then compute three--point amplitudes for the scattering of massless and massive spin--2 closed string states and reveal the double copy structure of the latter. With these results being finite in the string scale,  we are further able to reproduce the cubic Lagrangian of ghost--free bimetric theory around flat spacetime for bulk massive spin--2 states originating in products of vectors of extended brane supersymmetry.}

\begin{document}

\begin{flushright}
\hfill{LMU--ASC 02/23}

\hfill{MPP--2023--4}
\end{flushright}
\maketitle
\flushbottom

\section{Introduction} \label{intro}

As is well known, any string spectrum comprises infinite towers of \textit{physical} states, organised along distinct levels of increasing mass: their mass--squared is given by 
\beq
M^2_N = N\, \frac{1}{\alpha'}\,,
\eeq
where $N$ is an integer that grows with the level and $\alpha'$ the string scale. The seemingly discrete spectrum thus consists of infinitely many states that are \textit{all} relevant at $\alpha'$, with a mass gap essentially of the same order as the mass of any state. The level construction can proceed relatively most simply in the light--cone formulation, see for example \cite{Green:1987sp, Blumenhagen:2013fgp}. With these considerations holding both before and after the compactifications to four dimensions that we consider in this work, we denote with $D$ the generic number of total spacetime dimensions and proceed to highlight two universal properties of string states:
\begin{enumerate}
\item \label{first} they are viewed as \textit{mass eigenstates}, namely have on--shell mass, and 
\item \label{second} they correspond by construction to irreducible representations of the respective little group, that is ultimately $SO(D-2)$ and $SO(D-1)$ for massless and massive states respectively. A generic state's polarisation tensor is thus transverse and traceless. \end{enumerate}

Among these states, singlets, symmetric, antisymmetric as well as mixed--symmetry tensors of arbitrarily high ranks can be found, with the vast majority of string spectra consisting of states of the last kind. A practical way of visualising collectively string states is via 
the Regge trajectories they fall into, namely their distribution along straight lines of spin increasing linearly to their mass--squared, with spin here referring to the number of columns of the Young tableaux of their respective irreducible representations. It is further straightforward to observe that properties \ref{first} and \ref{second} take the form of the d'Alembert or the Klein--Gordon equation and of transversality and tracelessness constraints, as written in momentum space, respectively; we will collectively be referring to this set of equations as the on--shell conditions. It may thus come as enticing to associate each on--shell string state with a \textit{1--particle state}, in a spirit similar to that of Bargmann and Wigner, as comprehensively reviewed in \cite{Bekaert:2006py} in the context of field theory.
 
Physical string states can further be treated as external states of string scattering amplitudes, to compute which there exists a \textit{covariant} method which utilises (on--shell) vertex operators, namely suitable conformal fields of the two--dimensional CFT of the string worldsheet \cite{Weinberg:1985tv, Friedan:1985ey, Friedan:1985ge}. In particular, via the operator--state correspondence, any on--shell string state can be thought of, in a ``1--1'' fashion, as being created by a vertex operator, that captures its specific spin and mass properties, at an arbitrary point on the worldsheet. A (tree--level) scattering amplitude is then the expectation value of the product of the vertex operators that create its external states, integrated over all their possible insertion points on the worldsheet. Importantly, a long time ago it was demonstrated at tree--level \cite{Kawai:1985xq} that closed string amplitudes are nothing other than weighted sums of products of open string amplitudes, with the origin of such relations, known as Kawai--Lewellen--Tye (KLT) relations, lying in the property of closed string spectra of being tensor products of open string spectra. More recently, these results have been extended to one--loop \cite{Stieberger:2021daa, Stieberger:2022lss}.

A few decades after the discovery of KLT,  the radical formulation of gravitational amplitudes as products of Yang--Mills amplitudes unraveled gravity's double copy structure  \cite{Bern:2010ue}, thereby consolidating earlier evidence for such a link already implied by an $\alpha' \rightarrow 0$ limit of KLT relations \cite{Bern:1993wt, Bern:1998ug}. The double copy structure is based 
 on a duality between color and kinematics derived from the (conjectural) existence of relations between partial gluon subamplitudes \cite{Bern:2008qj}. The general proof of such tree--level BCJ amplitude identities has  subsequently been presented by using string theory and the power of world--sheet monodromy properties \cite{Stieberger:2009hq,Bjerrum-Bohr:2009ulz}.
The double copy structure has meanwhile not only seeded the classification and organisation of a very large web of field theories, for a panoramic review of which we refer the reader to \cite{Bern:2019prr}, but has further been employed as a greatly simplifying tool in regard to the study of the UV properties of supergravity beyond tree--level \cite{Bern:2007hh, Bern:2012cd, Bern:2013uka, Bern:2014sna, Bern:2018jmv}. Moreover, it has led to the construction of novel and powerful computational methods in regard to the perturbative treatment of the relativistic two--body problem, already to the extent of complementing and/or improving the numerical relativity modelling of the inspiral phase of spinless binary black hole systems \cite{Bjerrum-Bohr:2018xdl, Bern:2019nnu, Bern:2019crd, Bern:2021dqo}, today ever so timely in this era of gravitational--wave astronomy. For recent reviews of the matter including a complete list of references, as well as the value of the double copy towards increasing the precision of waveform models hence maximising gravitational--wave detection potential, the reader is referred to \cite{Buonanno:2022pgc, Adamo:2022dcm}.

In most of the above considerations, the graviton is assumed to be \textit{massless}. In the string side, it is identified with the symmetric (transverse and traceless) representation of the massless level of closed string spectra, since the interactions of the state in question perturbatively match those of the graviton of general relativity \cite{Scherk:1974ca, Green:1982sw, Gross:1986mw}; yet it is clear that both closed and open string spectra contain a profusion of massive spin--2 states. In the field side, the construction of healthy massive gravity as a modification of general relativity by a non--derivative potential came as a major breakthrough \cite{deRham:2010ik, deRham:2010kj} after decades of being widely considered as impossible \cite{Boulware:1972yco} beyond flat spacetime \cite{Pauli:1939xp, Fierz:1939ix}. A subsequent development inspired by the massive gravity potential was the formulation of bimetric theory \cite{Hassan:2011tf, Hassan:2011zd}, which around maximally symmetric backgrounds propagates, \textit{in addition to} the standard massless graviton, a massive spin-2 state \cite{Hassan:2011zd, Hassan:2012wr, Babichev:2016bxi, Babichev:2016hir} that has been put forward as a viable dark matter candidate \cite{Aoki:2016zgp, Babichev:2016bxi, Babichev:2016hir,Blanchet:2015sra}. For systematic reviews of massive gravity and bimetric theory the reader is referred to \cite{deRham:2014zqa, Hinterbichler:2011tt, Schmidt-May:2015vnx}.

More recently, the investigation of a double copy for massive gravity by means of Yang--Mills fields with Proca--like mass terms has demonstrated that, unlike the double copy of standard massless gravity, issues of locality arise beyond four--point (tree--level) interactions in the gravitational side \cite{Momeni:2020vvr, Johnson:2020pny}; this double copy has nevertheless been shown to be supersymmetrisable at the cubic level \cite{Engelbrecht:2022aao}. In the string side, it is worth noting that the original formulation of KLT relations, which yield a direct proof of double copy structure through world--sheet techniques, is independent of the mass level of the amplitudes' external states, although explicit examples involving solely massless states were given at the time \cite{Kawai:1985xq}; in the meantime, studies of the scattering of massive string states have either focused on lower--spin interactions or on leading Regge trajectories, starting with the seminal work \cite{Sagnotti:2010at}, followed by \cite{Feng:2010yx, Bianchi:2010es, Bianchi:2011se, Bianchi:2015yta, Bianchi:2015lnw, Lust:2021jps, Cangemi:2022abk}. In the field side, a modelling of massive higher spins has only very recently been achieved \cite{Ochirov:2022nqz} and, interestingly, a massive higher--spin symmetry has been shown to constrain Kerr amplitudes \cite{Cangemi:2022bew}. Besides a new and promising method involving both string and field theory tools has been devised to compute tree--level string amplitudes of one massive and arbitrarily many massless string states \cite{Guillen:2021mwp}, in the context of twisted strings \cite{Hohm:2013jaa, Huang:2016bdd, LipinskiJusinskas:2019cej}, where the worldsheet CFT tools can still be utilised as in string theories, but the spectrum is finite, more similarly to field theories. 

In continuation of our previous paper on massive spin--$2$ amplitudes and bimetric theory \cite{Lust:2021jps}, it is the scope of this work to probe the largely unexplored massive string states by delving into \textit{subleading} Regge trajectories. In particular, we consider massive brane vectors of such trajectories, that arise in compactifications to four spacetime dimensions, and construct their tensor products, namely massive spin--2 bulk states. We then proceed to investigate the self--interactions of the latter, as well as their interactions with the graviton, by calculating the respective $3$--point string scattering amplitudes. Despite the fact that a complete Lagrangian description of massive string states is deprived of meaning for reasons highlighted in the beginning of the introduction, with our $3$--point results being finite in $\alpha'$, we are able to extract ``effective'' Lagrangian cubic vertices and compare them with ghost--free bimetric theory as expanded to cubic order around Minkowski backgrounds. We find that, for bulk states originating in multiplets of $\mathcal{N}=4,8$ supersymmetry that also accommodate higher spins, an \textit{on--shell} match subject to an allowed tuning of the bimetric parameters is possible. Interestingly, since by construction the bulk spin--$2$ states in question transform nontrivially under the respective $R$--symmetry, while the spin--$2$ states of bimetric theory carry no charges, we select the singlet part of the former \textit{by hand}, thereby not preserving the full bulk supersymmetry in order to achieve this match. Crucially, in our previous work \cite{Lust:2021jps}, we showed that for the lowest--lying massive spin--2 state of compactified open strings, which belongs to a \textit{leading} Regge trajectory, such a match with bimetric theory is not possible. All our statements are to be taken strictly at the level of cubic interactions; going beyond three--point level is conceptually possible but tedious in string theory. In a different context, using twisted intersection theory, higher--point amplitudes and their underlying double copy structure have recently been presented for higher derivative gravity and certain corrections of the latter from partially massless bimetric gravity \cite{Mazloumi:2022nvi}. For holographic realisations of AdS massive gravity we also refer the reader to \cite{Porrati:2003sa, Kiritsis:2006hy, Aharony:2006hz, Kiritsis:2008at, Bachas:2017rch, Bachas:2018zmb, Bachas:2019rfq}.

The structure of this paper is as follows. In section \ref{probe} we begin by reviewing the vertex operator formulation of bosonic strings and of the superstring and proceed to scan the open string spectrum, after compactifications, for massive vectors with a structure as close to that of the massless vector as possible. These are associated with internal Kac--Moody currents and belong to multiplets of $\mathcal{N}=1,2,4$ brane supersymmetry. We then construct massive spin--2 closed string states, which fall into multiplets of $\mathcal{N}=2,4,8$ supersymmetry respectively, since the amount of brane supersymmetry preserved after compactifications is half that of the bulk, as is well known. Next, in section \ref{amplitudes} we compute $3$--point scattering amplitudes with these states and/or the graviton as external states by employing their double copy property in terms of the amplitudes of the respective vectors, which mirrors the standard massless double copy. Finally, in \ref{vertices} we use the amplitudes to extract cubic vertices which we compare with ghost--free bimetric theory. We also compare with the bosonic string at every step. Appendix \ref{conv_app} contains an illustration of the manifestation of Cardy's CFT trick in the string mass spectrum and appendix \ref{app_ope} a set of OPEs that we use throughout this work.

\ \\
\textbf{Conventions}. We use the mostly plus metric signature and restrict ourselves to a flat string background, in both the critical and four dimensions. Spacetime vector indices are denoted by letters around the middle of the Latin ($m,n,\dots\,$) and of the Greek ($\mu, \nu,\dots\,$) alphabet in the critical and in four dimensions respectively, such that $\eta^{mn}$ and $\eta^{\mu \nu}$ stand for the corresponding Minkowski metrics. Throughout the amplitudes section, we systematically omit overall unphysical numerical prefactors. In all our considerations, D--branes are assumed to be spacetime--filling. 

\section{Probing the string spectrum} \label{probe}

As is well--known, while in both the closed and the open string the worldsheet manifold is two--dimensional, it is only in the latter case that it has a boundary; as is customary, we will refer to the relevant CFTs as the ``bulk'' and the boundary CFT respectively, the corresponding systematics of each of which we gather in appendix \ref{conv_app}. In both cases, we can essentially think of a variable $z\in \mathbb{C}$, with $z_{12}\equiv z_1-z_2$, as parametrising the worldsheet coordinate. For the open string spectrum, the relevant worldsheet topology is that of a two--dimensional disk $D_2$, or equivalently, the upper half--plane (UHP) with the real axis as its boundary. For the closed string spectrum, it is a sphere $S^2$. Open and closed on--shell string states, of momentum $p^m$ and $k^m$ respectively, at mass level $n$ are thought of as being created by the vertex operators $V_n(z,p)$ and $V_n(z,\ov{z},k)$ at the points $z$ and $(z,\ov{z})$ of the respective worldsheet correspondingly. These are spacetime \textit{scalars} and also conformal fields which satisfy \cite{Weinberg:1985tv, Friedan:1985ey, Friedan:1985ge}
\beq \label{physical}
[Q, V]=\textrm{total derivative} \,,
\eeq
where $Q$ is the (worldsheet) nilpotent BRST charge of the string theory in question; (\ref{physical}) is clearly reminiscent of the criteria used to select physical states in quantum field theories. We adhere mostly to the conventions of \cite{Blumenhagen:2013fgp}, with the distinction that we reinstate $\alpha'$ in all formulae; this step is essential, as the low--energy limit within string theory, at least for \textit{massless} external states, is broadly understood as $\alpha' \rightarrow 0$ \cite{Neveu:1971mu, Gervais:1972tr, Scherk:1971xy, Nakanishi:1972zp, Frampton:1973sx, Scherk:1974ca, Green:1982sw, Gross:1986mw}.

\subsection{The critical open string}

As will become clear later, it is instructive to begin with a scan of the open bosonic string's spectrum. The mass--squared of its $n$th mass--level is given by
\beq \label{Regge_open_b}
M_{n,\,\textrm{open}}^2 = (n-1)\, \frac{1}{\alpha'} \quad , \quad n\, \in \mathbb{N}_0 \,.
\eeq
The worldsheet CFTs involve here the following holomorphic ingredients:
\begin{itemize}
\item the spacetime vectors $X^m(z)$ as embedding coordinates, which we will be referring to as ``matter'' fields as is standard. The corresponding energy--momentum tensor takes the form
\beq\label{boson_EM}
T^{X}(z)= -\frac{1}{4\alpha'} \, \partial X \partial X\,(z)\,,
\eeq
where summation over the spacetime index $m$ is implied, while their two--point function reads
\beq \label{boson_two}
\langle X^m(z_1) X^n(z_2) \rangle_{D_2} = - 2 \alpha' \eta^{m n} \, \ln z_{12} \,.
\eeq
\item the worldsheet conformal (and anti--commuting) $b(z),c(z)$ ghosts. The corresponding energy--momentum tensor takes the form
\beq \label{ghost_EM_b}
T^{b,c}(z)  =-2\,b\partial c(z)- (\partial b)c(z)\,,
\eeq
while their two-- and three--point functions read respectively
\beq \label{ghost_two_three}
 \langle c(z_1)b(z_2) \rangle = \frac{1}{z_{12}} \quad , \quad  \langle  c(z_1)c(z_2)c(z_3) \rangle = z_{12}z_{13}z_{23}\,.
\eeq
\end{itemize}
The number of spacetime dimensions, or equivalently of ``matter'' fields, is fixed to $26$ by requiring the cancellation of the total conformal anomaly on the worldsheet: the value of the central charge of the ghost CFT is $-24$, so a zero net central charge is achieved for a ``matter'' CFT with central charge equal to $24$. Notice that these two CFTs can be thought of as \textit{decoupled} from each other, since ghosts and ``matter'' do not interact. 

The total momentum and total angular momentum of the bosonic string are respectively given by
\beq \label{currents_b}
P_m =\frac{1}{2\alpha'}\, \oint \frac{\md z}{2\pi i} \, i \partial X_m \quad , \quad J^{mn} =   \frac{1}{2\alpha'}\,  \oint \frac{\md z}{2\pi i} \, iX^{[m} i \partial X^{n]} \,.
\eeq
Moreover, the BRST charge, which we denote with $Q_0$, takes the form
\beq \label{charge_b}
\begin{array}{ccl}
Q_0&=&\oint \frac{\md z}{2\pi i}\Big[c \big(T^{X}+\frac{1}{2}T^{b,c}\big) \Big] =\oint \frac{\md z}{2\pi i} \,c \, \Big[T^X+(\partial c) b   \Big]\,,
\end{array}
\eeq
substituting which in (\ref{physical}) yields
\beq \label{physical2}
h_V\, ( \partial c) V(z) + c\,\partial V(z) \overset{!}{=} \textrm{tot. deriv.} \,,
\eeq
where by $h_V$ we denote the conformal weight of a generic open string vertex operator $V(z)$. (\ref{physical2}) then implies that \textit{all} physical such operators must have weight
\beq \label{physical3}
h_V=1\,.
\eeq

A vertex operator $V_n(z,p) $, that creates the \textit{full} set of on--shell states at the $n$--th level, contains then the following building blocks:
\begin{enumerate}
\item naturally the momentum eigenstate $e^{ipX(z)}$, which is a primary field of weight $\alpha' p^2=1-n$, where we have used (\ref{Regge_open_b}), and
\item a polynomial $\mathcal{V}_n$ in $\partial X^m (z)$, namely a composite conformal operator that, using (\ref{physical3}) and the fact that conformal weights are additive, must have weight
\beq
h_{\mathcal{V}_n}=n\,.
\eeq
\end{enumerate}
Quite generally then
\beq \label{genVO_b}
V_n(z,p) = g_{\textrm{o}} \, T^a \,   \mathcal{V}_n\Big(  \partial X^m(z),p \Big) \, e^{ipX(z)} \quad , \quad p^2=-\frac{n-1}{\alpha'}\,,
\eeq
where $g_{\textrm{o}}$ is the open string coupling and $T^a$ the brane gauge group generator, with normalisation such that
\beq \label{nonab_alg}
\begin{array}{ccl}
\Tr\big(T^a T^b \big)= \frac{1}{2} \,\delta^{ab} \quad , \quad [T^a, T^b]=if^{abc} \,T^c\,.
\end{array}
\eeq
Notice that $T^a$ is introduced à la Chan--Paton \textit{by hand} in (\ref{genVO_b}). $\mathcal{V}_n$ is thus a linear combination of all possible spacetime scalars of weight $n$ constructed out of $\partial X^m$, to formulate which one needs to introduce a priori arbitrary tensors that contract $\partial X^m$ and its powers. For a given $n$, (\ref{physical}) then not only forces (\ref{physical3}), but also algebraic constraints on these tensors, whose solutions are on--shell conditions such as transversality and tracelessness, as previously referred to. In this way, $V_n$ splits into the vertex operators that create all physical states at the given $n$, which we will denote with $V_{n,\textrm{open}}$. It should be clear that this procedure is performed on a level by level basis and may become cumbersome for very heavy states, as the number of states grows exponentially with $\sqrt{n}$, which is reflected in the growth of the weight of $\mathcal{V}_n$. An alternative is the DDF construction \cite{DelGiudice:1971yjh}, in which the vertex operator for an arbitrarily massive string state can be built by means of computing the amplitude corresponding to the scattering of arbitrarily many photons off a tachyon.

We are now ready to review the first few levels of the open bosonic spectrum:
\begin{itemize}
\item level $0$: it contains a single scalar tachyon propagating $1$ dof, with vertex operator
\beq \label{tachyon_open_b}
V_{\textrm{t,o}}(z,p)=g_{\textrm{o}}\, T^a \,  e^{ipX(z)} \quad , \quad p^2=\frac{1}{\alpha'}\,.
\eeq

\item level $1$: it contains a single massless vector propagating $24$ dof, with vertex operator
\begin{equation} \label{gaugevo_open_b}
V_A (z,\epsilon,p) = \frac{g_{\textrm{o}}}{\sqrt{2\alpha'}} \, T^a\, \epsilon_m \partial X^m(z) \, e^{ipX(z)}\ \quad , \quad p^2=0 \quad , \quad p \cdot \epsilon = 0\,,
\end{equation}
where $\epsilon_m$ is a (constrained) vector of $SO(1,25)$.
\item level $2$: it contains a single massive spin--2 state propagating $324$ dof, with vertex operator \cite{Bianchi:2015yta}
\beq \label{vertop_open_b}
\begin{array}{ccl}
V_B (z,B,p)= \frac{g_{\textrm{o}}}{2\alpha'} \, T^a\, B_{mn} \,i \partial X^m (z) i\partial X^n(z) \ e^{ipX(z)}\,,
\crbig
p^2 = - \frac{1}{\alpha'}  \quad , \quad  B_{[mn]}=0 \quad , \quad p^m B_{mn} = 0\quad , \quad B_m^m=0 \,,
\end{array}
\eeq
where $B_{mn}$ is a (constrained) rank--2 tensor of $SO(1,25)$. Notice that (\ref{vertop_open_b}) is the ``\textit{lowest}'' lying massive ``graviton'' of the open bosonic string spectrum in $26$ dimensions.
\end{itemize} 

Turning next to the superstring, the mass--squared of the $n$th mass--level of the open superstring spectrum is given by
\beq \label{Regge_open}
M_{n,\,\textrm{open}}^2 = n\, \frac{1}{\alpha'} \quad , \quad n\, \in \mathbb{N}_0 \,.
\eeq
In the NS sector of the RNS formalism \cite{Neveu:1971rx,Ramond:1971gb}, the worldsheet CFTs now also involve the superpartners of the holomorphic ingredients of those of the bosonic string, namely:
\begin{itemize}
\item the spacetime vectors $X^m(z)$ and $\psi^m(z)$ as embedding coordinates, which we will be referring to as ``matter'' fields as is customary. These are superpartners under $\mathcal{N}=1$ (worldsheet) supersymmetry, that is generated by the energy--momentum tensor $T(z)$ and supercurrent $T_{\textrm{F}}(z)$ given by the composite operators
\beq \label{supc_m}
\begin{array}{ccl}
T^{X,\psi}(z)&=& T^{X}(z)+T^{\psi}(z) = -\frac{1}{4\alpha'} \, \partial X \partial X\,(z)-\frac{1}{2}\, \psi \partial \psi \,(z)
\crbig
T^{X,\psi}_{\textrm{F}}(z)&=& \frac{i}{2\sqrt{2\alpha'}}\, \psi \partial X \, (z)\,.
\end{array}
\eeq
The relevant two--point functions are (\ref{boson_two}) and 
\beq \label{open_opes_ten}
\begin{array}{ccl}
 \langle \psi^m(z_1) \psi^n(z_2)  \rangle = \frac{\eta^{mn}}{z_{12}} \,.
 \end{array}
\eeq
We suppress the (worldsheet) spinor index of $\psi^m$ and $T_{\textrm{F}}\,$, as there is no confusion. 
\item the worldsheet conformal (and anti--commuting) $b(z),c(z)$ and superconformal (and commuting) $\beta(z),\gamma(z)$ ghosts. The pairs $\beta, b$ and $c,\gamma$ are superpartners under $\mathcal{N}=1$ (worldsheet) supersymmetry, with their contributions $T_{\textrm{gh}}$ and $T_{\textrm{F},\textrm{gh}}$ to the energy--momentum tensor and supercurrent respectively given by
\beq \label{supc_g}
\begin{array}{ccl}
T^{\textrm{gh}}(z)&=& T^{b,c}(z) + T^{\beta, \gamma}(z) =-2\,b\partial c(z)- (\partial b)c(z) -\frac{3}{2}\,\beta \partial \gamma(z) -\frac{1}{2}\,(\partial \beta ) \gamma(z)
\crbig
T_{\textrm{F}}^{\textrm{gh}}(z)&=& \frac{1}{2}\, b\gamma(z) - (\partial \beta) c(z) -\frac{3}{2} \, \beta \partial c(z)  \,.
\end{array}
\eeq
The relevant two-- and three--point functions now are (\ref{ghost_two_three}) and
\beq  \label{open_opes_ten_gh}
\begin{array}{ccl}
\langle \gamma(z_1) \beta(z_2) \rangle = \frac{1}{z_{12}}\,.
\end{array}
\eeq

\end{itemize}
\begin{table}
\centering 
\renewcommand{\arraystretch}{1.5}
  \begin{tabular}{ c || c | c | c  | c | c | c | c }
   object & $\alpha'$ & $X^m$, $c$ & $\gamma$ & $Q$ & $\psi^m$ & $\beta$, $T_F$ & $b$, $T$ \\ \hline
   mass dimension & $-2$ & $-1$ & $-1/2$ & $0$ & $1/2$ & $3/2$ & $2$ 
  \end{tabular}
\renewcommand{\arraystretch}{1}
\caption{Mass dimensions of parameters, fields and composite operators} \label{massd}
\end{table}
The number of spacetime dimensions, or equivalently of ``matter'' superpartners, is now fixed to ten by requiring the cancellation of the total conformal anomaly on the worldsheet: the value of the central charge of the ghost and superghost CFTs is respectively $-26$ and $11$, so a zero net central charge is achieved for a ``matter'' CFT with central charge equal to $15$, or equivalently ten pairs of superpartners $X^m, \psi^m$. Notice that these three CFTs can again be thought of as \textit{decoupled} from each other, since ghosts and ``matter'' do not interact with each other.

The total momentum and total angular momentum of the superstring are respectively given by
\beq
P_m =\frac{1}{2\alpha'}\, \oint \frac{\md z}{2\pi i} \, i \partial X_m \quad , \quad J^{mn} = \oint \frac{\md z}{2\pi i} \, \Big\{ \frac{1}{2\alpha'}\, iX^{[m} i \partial X^{n]} + \psi^{[m} \psi^{n]} \Big\}\,.
\eeq
Moreover, $Q$ now takes the form
\beq
\begin{array}{ccl}
Q&=&\oint \frac{\md z}{2\pi i}\Big[c \big(T^{X,\psi}+\frac{1}{2}T^{\textrm{gh}}\big) -\gamma \big(T^{X,\psi}_{\textrm{F}}+\frac{1}{2}T_{\textrm{F}}^{\textrm{gh}}\big)\Big] = Q_0 + Q_1 + Q_2 \,,
\end{array}
\eeq
where
\beq \label{brst_compo}
\begin{array}{ccl}
Q_0 &=& \oint \frac{\md z}{2\pi i} \, c \, \Big[T^{X,\psi}+T^{\beta,\gamma} +(\partial c) b \Big]
\crbig
Q_1 &=& -\frac{i}{2\sqrt{2\alpha'}} \oint \frac{\md z}{2\pi i} \,e^{-\chi} e^{\phi} \, \psi \partial X
\crbig
Q_2 &=&\frac{1}{4} \oint \frac{\md z}{2\pi i} \, b \, e^{2\phi} e^{-2\chi}\,,
\end{array}
\eeq
with the $(\beta,\gamma)$ system appearing in its bosonised form for simplicity
\beq
\beta =   e^{-\phi } e^\chi \partial \chi \quad , \quad \gamma = e^{-\chi} e^\phi\,,
\eeq
where $\phi$ and $\chi$ are chiral bosons with two--point functions that read
\beq
\begin{array}{ccl}
\langle \chi(z_1) \chi(z_2) \rangle = -\langle \phi(z_1)\phi(z_2) \rangle = \ln z_{12}\,.
\end{array}
\eeq
Notice that, up to fermions and superghosts, $Q_0$ is equal to the total BRST charge (\ref{charge_b}) of the bosonic string. For completeness, we list the mass dimensions of a selection the aforementioned fields and operators in table \ref{massd} for convenience, where $T$ and $T_F$ stand for the energy--momentum tensor and the supercurrent of either ``matter'' or ghosts respectively. Interestingly, for $\psi^m$ and the constituent fields of the ghost systems, as well as for the supercurrent multiplets, the mass dimensions are \textit{equal} to the respective conformal weights.

\begin{table}
\centering 
\renewcommand{\arraystretch}{1.5}
  \begin{tabular}{ c || c | c | c  | c  }
   operator & $\partial X^m(z)$ &  $\psi^m(z)$ & $e^{q\phi(z)}$ & $e^{ipX(z)}$ \\ \hline
   weight & $1$ & $1/2$  & $-\frac{1}{2}q(q+2)$ & $-n$   
  \end{tabular}
\renewcommand{\arraystretch}{1}
\caption{Conformal weights of the primary fields entering open string vertex operators} \label{confd}
\end{table}

A vertex operator $V_n^{(q)}(z,p) $, that creates the \textit{full} set of on--shell states at the $n$--th level, contains then the following building blocks, whose conformal weights we list in table \ref{confd}:
\begin{enumerate}
\item naturally the momentum eigenstate $e^{ipX(z)}$,
\item the object $e^{q\phi (z)}$, with $q$ being its charge under the anomalous $U(1)$ symmetry of the $(\beta, \gamma)$ system, usually referred to as the ``ghost picture'' of $V_n^{(q)}(z,p) $. This object captures the property of superstring states of being associated with an infinite degeneracy along the Bose sea of the  $(\beta, \gamma)$ system. In the NS sector on which we focus, $q \in \mathbb{Z}$.
\item a polynomial $\mathcal{V}_n$ in $\partial X^m (z)$ and $\psi^m(z)$, namely a composite conformal operator. For a given $q$, (\ref{physical}) now implies separately that
\beq \label{physical_superstring}
[Q_0, V(z)]=[Q_1,V(z)]=[Q_2,V(z)]=0\,,
\eeq
the important consequence of the first of which being again (\ref{physical3}). Using table \ref{confd}, $\mathcal{V}_n$ must then have weight
\beq \label{weightV}
h_{\mathcal{V}_n} = 1+\frac{1}{2}q(q+2)+n 
\eeq
for a given set of $(q,n)$. 
\end{enumerate}
Quite generally then
\beq \label{genVO}
V_n^{(q)}(z,p) = g_{\textrm{o}} \, T^a \,  e^{q\phi(z)}\,  \mathcal{V}_n\Big(  \partial X^m(z), \psi^m(z) , p \Big) \, e^{ipX(z)} \quad , \quad p^2=-\frac{n}{\alpha'}\,,
\eeq
$\mathcal{V}_n$ is thus a linear combination of all possible spacetime scalars with weight given by (\ref{weightV}) constructed out of $\partial X^m$ and $\psi^m$. Notice that for a given picture $q$, as the level $n$ increases, $h_{\mathcal{V}_n}$ also increases and so does the number of possible operators that contribute to $\mathcal{V}_n$. For a given level $n$, the ghost picture in which the fewest possible operators are required to build $\mathcal{V}_n$ is called ``canonical'', and it is $-1$ in the NS sector. Due to (\ref{physical_superstring}), $V_n^{(q)}$ splits into the vertex operators that create all physical states for a given set $(q,n)$, which we will denote with $V^{(q)}_{n,\textrm{open}}$, similarly to the bosonic case.  

We are now ready to review the first few levels of the open superstring spectrum in the canonical ghost picture:
\begin{itemize}
\item level $0$: $\textbf{8}_{\textrm{B}}+\textbf{8}_{\textrm{F}}$. The NS sector contains a single massless vector with vertex operator
\begin{equation} \label{gaugevo}
V_A^{(-1)}(z,\epsilon,p) = g_{\textrm{o}} \, T^a\, e^{-\phi(z)} \, \epsilon_m \psi^m(z) \, e^{ipX(z)} \quad, \quad p^2=0 \quad , \quad p \cdot \epsilon = 0\,,
\end{equation}
where $e_m$ is a (constrained) vector of $SO(1,9)$.
\item level $1$: $\textbf{128}_{\textrm{B}}+\textbf{128}_{\textrm{F}}$. The NS sector contains two states \cite{Tanii:1986ug, Tanii:1987bk, Koh:1987hm}, a spin--$(1,1,1)$ which propagates $84$ dof
\beq \label{vertop1}
\begin{array}{ccl}
V_E^{(-1)}(z,E,p)= g_{\textrm{o}} \, T^a\,\ e^{-\phi(z)}\, E_{m n p} \, \psi^m \psi^n \psi^p(z) \ e^{ipX(z)} 
\crbig
p^2=-\frac{1}{\alpha'} \quad , \quad p^m E_{mnp}=0\,,
\end{array}
\eeq
where $E_{mnp}$ is a completely antisymmetric (constrained) rank--3 tensor of $SO(10)$ and a spin--$2$ which propagates $44$ dof
\beq \label{vertop}
\begin{array}{ccl}
V_B^{(-1)}(z,B,p)= \frac{g_{\textrm{o}}}{\sqrt{2\alpha'}}\ T^a\ e^{-\phi(z)}\, B_{mn} \,i \partial X^m (z) \psi^n(z) \ e^{ipX(z)}\,,
\crbig
p^2 = - \frac{1}{\alpha'} \quad , \quad B_{[mn]}=0 \quad , \quad p^m B_{mn} = 0\quad , \quad B_m^m=0 \,,
\end{array}
\eeq
where $B_{mn}$ is a (constrained) rank--2 tensor of $SO(1,9)$. Notice that the latter is the \textit{lowest} lying massive ``graviton'' of the open superstring spectrum in ten dimensions.
\end{itemize} 

Finally, let us note that the open superstring's levels are organised in multiplets of (spacetime) $\mathcal{N}=1$, $D=10$ supersymmetry, where spacetime--filling D--branes are assumed. For completeness, the $\mathcal{N}=1$, $D=10$ algebra in the ``ghost--neutral'' picture reads
\beq \label{susy10d}
\begin{array}{ccl}
\{ \mathcal{Q}_\alpha^{(+1/2)}, \mathcal{Q}_\beta^{(-1/2)} \} = (\gamma^m C)_{\alpha \beta} \, P_m\,,
\end{array}
\eeq
where the supersymmetry generator $\mathcal{Q}_\alpha\,$ is given, in the $(-1/2)$ and in the $(+1/2)$ ghost picture respectively, by
\beq
Q_\alpha^{(-1/2)} =  \frac{1}{\alpha'^{1/4}}\, \oint \frac{\md z}{2\pi i} \, S_\alpha e^{-\phi/2} \quad , \quad Q_\alpha^{(+1/2)} = \frac{1}{2\alpha'^{3/4}}\, \oint \frac{\md z}{2\pi i} \, i \partial X_m \, (\gamma^m S)_\alpha \, e^{\phi/2}\,.
\eeq
In the above, $S_\alpha(z)\,, \alpha=1,\dots,16\,$ is a (left--handed) spinor of the Lorentz group $SO(1,9)$, has conformal dimension $5/8$ and is used to build states in the R sector; importantly, its OPE with $\psi^m$ is \textit{not} regular:
\beq \label{branch_cut}
\begin{array}{ccl}
\psi^m(z) S_\alpha(w) \sim\displaystyle{ \frac{\gamma^m_{\alpha \dot{\beta}}}{\sqrt{2} \, (z-w)^{1/2}}\, S^{\dot{\beta}}(w)\,.}
\end{array}
\eeq

\subsection{Compactifications to four dimensions} \label{section_comp_open}

We now turn to superstring compactifications from ten to four space--time dimensions following the symmetry breaking pattern
\beq \label{sym_break}
 SO(1,9)\rightarrow SO(1,3)\times SO(6)\,. 
\eeq 
The $D=10$ ``matter'' fields $\partial X^m,\psi^m$ and $S_\alpha$ give rise to the (universal) $D=4$ fields 
\beq \label{split}
 \partial X^\mu,  \psi^\mu \quad , \quad \mu = 0,\dots,3  \quad , \quad S_a , S^{\dot{a}}\quad , \quad a,\dot{a}=1,2
\eeq
so that $i \partial X^\mu$ and $\psi^\mu$ are vectors, while $S_a$ a spinor of $SO(1,3)$.
In addition, we have the internal Ramond fields 
\beq\label{split1}
(\Sigma^I \,,\,  \ov{\Sigma}_{I})\quad , \quad I=1,\ldots{\cal N}
\eeq
falling into representations of the $R$--symmetry. The index $I$ essentially enumerates the surviving spacetime supersymmetries in four space--time dimensions. 
While the fields $X^\mu,\psi^\mu$  inherit their weights of their ancestors $i \partial X^m$ and $\psi^m$,  the weights of $S_a$ and $\Sigma^I$ are $1/4$ and $3/8$, respectively. 
Additional internal fields 
\beq\label{fieldsI}
\partial Z^M,\Psi^M\quad , \quad M=1,\ldots,6
\eeq may exist for extended supersymmetries. For the  fields (\ref{split}) and (\ref{fieldsI}), OPEs and correlators follow from their ancestors in the critical dimension as outlined in the previous section, for example 
\beq \label{open_opes_four}
\begin{array}{ccl}
\langle X^\mu(z_1) X^\nu(z_2) \rangle_{D_2} = - 2 \alpha' \eta^{\mu \nu} \, \ln z_{12} \quad , \quad  \langle \psi^\mu(z_1) \psi^\nu(z_2)  \rangle = \frac{\eta^{\mu \nu}}{z_{12}} 
\crbig
\langle Z^M(z_1) Z^N(z_2) \rangle_{D_2} = - 2 \alpha' \delta^{MN} \, \ln z_{12} \quad , \quad  \langle \Psi^M(z_1) \Psi^N(z_2)  \rangle = \frac{\delta^{MN}}{z_{12}} \,.
 \end{array}
\eeq
Due to the rule (\ref{sym_break}), it is straightforward to see that the CFTs of the internal fields decouple from the those of the ``matter'' fields in the sense that mixed OPEs are regular; this is a crucial property of which we are going to make repeated use in the following. Note, however, the former may still contribute to physical vertex operators in four dimensions, as they combine into primary fields, in a manner determined by the compactification in question; we will consider only the ones relevant for the present work on a case by case basis. Finally, the ghost and superghost CFTs also remain as are in the critical dimension, since all ghosts are spacetime scalars. 

In regard to the spectrum, we restrict ourselves to states that carry \textit{no} internal momenta, namely states for which 
\beq \label{no_leak}
p^m=(p^\mu,0)\,,
\eeq
so that the mass spacing remains intact and the mass spectrum (\ref{Regge_open}) is equally well valid in ten and in four dimensions; note that consequently in all momentum eigenstates in four dimensions, summation over the spacetime index $\mu$ is assumed. Moreover, it should be clear that the operator--state correspondence is valid both before and after compactifications, so that all formulae of the previous subsection can be employed upon using (\ref{split}); for example, the worldsheet supercurrent, given by the second of equations (\ref{supc_m}), generically splits as
\beq \label{sc_split}
T^{X,\psi}_{\textrm{F}}(z)= \frac{i}{2\sqrt{2\alpha'}}\, \psi \partial X \, (z)+ 
T_{\textrm{F,int}}(z) \  ,
\eeq
where summation over the $\mu$ indices is implied in the first term. Furthermore, the second term  $T_{\textrm{F, int}}$ accounts for the internal supercurrent to be specified in the sequel. The internal CFT has central charge equal to $9$. Depending on the compactification manifold, part or all of the spacetime supersymmetries in ten space--time dimensions are preserved on the brane worldvolume (and likewise in the bulk); this is essentially achieved by projecting out part or none of the internal fields $\Sigma^I$. Generically, internal fields combine to form currents $\mathcal{J}^{IJ}$ of conformal weight $1$, that arise in the OPE of $\Sigma^I$, $\ov{\Sigma}^{I}$ for ${\cal N}=2,4$ \cite{Banks:1987cy, Banks:1988yz} 
\beq
\begin{array}{ccl}
\Sigma^I(z) \ov{\Sigma}^J(w) &=&\displaystyle{ \frac{\delta^{IJ} }{(z-w)^{3/4}}\ \mathcal{I}+ (z-w)^{1/4}\, \mathcal{J}^{IJ}(w)+\dots\ ,}
\crbig
\Sigma^I(z) \Sigma^J(w) & \sim & (z-w)^{1/4} \, \psi^{IJ}(w)+\ldots\,,
\end{array}
\eeq
where $\mathcal{I}$ is the identity operator, $\psi^{IJ}$ is a conformal weight $1/2$ operator and the dots stand for regular terms. Likewise  for ${\cal N}=1$ we have \cite{Banks:1987cy}
\beq
\begin{array}{ccl}
\Sigma(z) \ov{\Sigma}(w) &=&\displaystyle{ \frac{1}{(z-w)^{3/4}}\ \mathcal{I}+\frac{\sqrt 3}{2}  (z-w)^{1/4}\, \mathcal{J}(w)+\dots\ ,}
\crbig
\Sigma(z) \Sigma(w) & \sim & (z-w)^{3/4} \, {\cal O}(w)+\ldots\ ,
\end{array}
\eeq
the dimension one current ${\cal J}$ and  some dimension $3/2$ operator ${\cal O}$ appear.


The currents ${\cal J}$ and $\mathcal{J}^{IJ}$ furnish a Kac--Moody algebra $\mathfrak{g}$ of level $1$ \cite{Banks:1987cy, Banks:1988yz} that has been shown to be a symmetry of the \textit{spectrum} \cite{Ferrara:1989ud}. In particular, for four--dimensional string compactifications with D-branes and  orientifolds (cf. \cite{Blumenhagen:2006ci} for a detailed account) one can distinguish the following cases:
\begin{enumerate}
\item $\mathcal{N}=1$, $\mathfrak{g}= \mathfrak{u}(1) $. This case can be achieved for example by Calabi-Yau orientifolds with D3/D7, D5/D9 or intersecting D6--branes. There exists one pair $\Sigma^+\equiv\Sigma$, $\Sigma^-\equiv \bar\Sigma$ that is associated with a $U(1)$ current $\mathcal{J}$ with Kac--Moody algebra
\beq \label{km_1}
\langle \mathcal{J}(z_1) \mathcal{J}(z_2) \rangle = \frac{1}{z_{12}^2}  \quad, \quad  \langle \mathcal{J}(z_1)\mathcal{J}(z_2)\mathcal{J}(z_3)  \rangle = 0\,.
\eeq
The internal supercurrent $T_{\textrm{F,int}}$ entering (\ref{sc_split}) splits into two components 
\beq \label{split_1}
T_{\textrm{F,int}} \equiv \frac{1}{\sqrt{2}}\big(T_{\textrm{F,int}}^+ + T_{\textrm{F,int}}^- \big)\,,
\eeq
of opposite $U(1)$ charge under $\mathcal{J}$ with
\beq \label{ope_current_superc}
 \mathcal{J}(z) \,T_{\textrm{F,int}}^{\pm} (w)  = \pm \frac{T_{\textrm{F,int}}^{\pm} (w)}{\sqrt{3} (z-w)} +  \mathcal{J}(w)\,T_{\textrm{F,int}}^{\pm} (w) + \dots
\eeq
thereby invoking the enhancement $\mathcal{N}=1 \rightarrow \mathcal{N}=2$ of \textit{worldsheet} supersymmetry \cite{Banks:1987cy, Hull:1985jv}.
Above the  dots stand for regular terms.
Eventually, one may write
\beq
\mathcal{J} (z)\equiv i \partial H (z) \quad , \quad  \Sigma=e^{i\tfrac{\sqrt3}{2}H}\quad , \quad  {\cal O}=e^{i\sqrt3H}
\eeq
where
$H$ is a free scalar with 2--point function
\beq \label{free_abelian}
\langle H(z) H(w) \rangle = - \ln (z-w) \,,
\eeq
such that an operator of the form $e^{iqH(z)}$ has charge $q$ under $\mathcal{J}$ and conformal weight $\frac{1}{2}q^2\,$.

\item $\mathcal{N}=2$, $\mathfrak{g}=\mathfrak{u}(1) \times \mathfrak{su}(2)  $. This can be achieved for example by $K3 \times T^2$ orientifolds with D3/D7 or D5/D9--branes. There exist two pairs $\Sigma^I,\bar\Sigma^I,\;I=1,2$ of spin fields which factorize and can be shown \cite{Banks:1988yz} to be associated with a $U(1)$ current $\mathcal{J}$ as well as an $SU(2)$ triplet of currents $\mathcal{J}^A$, $A=1,2,3$ with (level one) Kac--Moody algebra
\beq \label{km_2}
\begin{array}{ccl}
\mathcal{J}^A(z_1) \mathcal{J}^B(z_2) \sim \frac{ \delta^{AB}}{z_{12}^2} +  \frac{i\sqrt{2}\,\varepsilon^{ABC}}{z_{12}} \, \mathcal{J}^C(z_2)
\crbig
 \langle \mathcal{J}^A(z_1)\mathcal{J}^B(z_2)\mathcal{J}^C(z_3) \rangle = \frac{\varepsilon^{ABC}}{z_{12}z_{13}z_{23}}\,.
\end{array}
\eeq
The currents $\mathcal{J}$ and $\mathcal{J}^3$ can be written as
\beq
\mathcal{J} (z) \equiv i \partial H_s (z) \quad, \quad \mathcal{J}^3 (z) \equiv i \partial H_3 (z) \,,
\eeq
with $H_s$ and $H_3$ two free (and decoupled) scalars enjoying  (\ref{free_abelian}).
The internal supercurrent $T_{\textrm{F,int}}$ entering (\ref{sc_split}) splits in two components that anticommute as
\beq \label{split_2}
T_{\textrm{F,int}} = T_{\textrm{F,int}}^{c=3} + T_{\textrm{F,int}}^{c=6}\ ,
\eeq
where $T_{\textrm{F,int}}^{c=3}$ is a $U(1)$ doublet
carrying no $SU(2)$ charge and $T_{\textrm{F,int}}^{c=6}$ an $SU(2)$ doublet carrying no $U(1)$ charge.
More precisely, the $c=3$ part is given by one (internal) complex boson $Z$ (with $h=1$) and
a complex fermion $\Psi=e^{iH_s}$ as
\beq
T_{\textrm{F,int}}^{c=3}=\frac{1}{2\sqrt{2\alpha'}}\ (i\partial Z\,e^{-H_s}+i\partial \bar Z\,e^{H_s})\ .
\eeq
Furthermore, $T_{\textrm{F,int}}^{c=6}$ has two components of charge $\pm 1/\sqrt{2}$ under $\mathcal{J}^3$
\beq \label{sc_2_doublet}
\begin{array}{ccl}
T_{\textrm{F,int}}^{c=6} (z) =\displaystyle{\frac{1}{\sqrt{2}} \, \sum\limits_{i=1}^2 \lambda^i(z) \, g_i(z) \ ,}
\end{array}
\eeq
where $\lambda^i$ and $g_i$ have conformal weights $1/4$ and $5/4$ respectively and
\beq
\Sigma^i = \lambda^i \,e^{\frac{i}{2}H_s} \quad, \quad \lambda^{1,2} \equiv e^{\pm \frac{i}{\sqrt{2}}H_3}\,.
\eeq
Crucially, $ T_{\textrm{F,int}}^{c=3}$ and $g_i$ are \textit{decoupled} from $\mathcal{J}^A$ and $\lambda^i$ so, at least for the states and amplitudes we will consider in this work, knowledge of the OPE
\beq \label{nonabelian_sc_current}
\begin{array}{ccl}
\lambda^i(z) \mathcal{J}^A(w) = \frac{(\tau^A)^i_j}{\sqrt{2}(z-w)} \, \lambda^j(w) - \frac{1}{\sqrt{2}}(\tau^A)^i_j \, \partial \lambda^j(w) + \dots
\end{array}
\eeq
where the dots stand for regular terms and $\tau^A$ are the Pauli matrices of $SU(2)$, suffices, as we will show. For completeness, let us note that $\mathcal{J}$ and $\mathcal{J}^A$  invoke the enhancements $\mathcal{N}=1 \rightarrow \mathcal{N}=2$ and $\mathcal{N}=1 \rightarrow \mathcal{N}=4$ of \textit{worldsheet} supersymmetry respectively \cite{Banks:1988yz} .

\item $\mathcal{N}=4$, $\mathfrak{g}=\mathfrak{so}(6) \times \mathfrak{u}(1)^6 $. This is the situation of a maximally supersymmetric compactification on a generic $T^6$ torus with the six internal coordinates $Z^M(z),\;M=1,\ldots,6$. As a consequence all brane supersymmetries are preserved. The internal supercurrent $T_{\textrm{F,int}}$ entering (\ref{sc_split}) is given by 
\beq
T_{\textrm{F,int}}(z)=\frac{i}{2\sqrt{2\alpha'}} \sum_{M=1}^6\Psi^M\,\partial Z^M(z)\ .
\eeq
The $SO(6)\simeq SU(4)$ current takes the form given for example in \cite{Ferrara:1989ud}:
\beq \label{current_4}
\mathcal{J}^{MN} (z) = \frac{1}{\sqrt{2}} \,\Psi^M \Psi^N (z)\,.
\eeq
Using (\ref{current_4}) and the internal fermion OPE as deduced by (\ref{OPE_exp}), we calculate 
\beq \label{km_4}
\begin{array}{ccl}
\mathcal{J}_{MN}(z_1) \mathcal{J}^{KL}(z_2) \sim\displaystyle  \frac{ \delta_M^{[K} \delta_N^{L]} }{z_{12}^2} +  \frac{2\sqrt{2}}{z_{12}} \, \delta_{(M}^{[K} \mathcal{J}_{N)}^{L]}(z_2)\ ,
\crbig
 \langle \mathcal{J}^{MN}(z_1)\mathcal{J}_{KL}(z_2)\mathcal{J}^{PR}(z_3) \rangle =\displaystyle \frac{\sqrt{2} }{z_{12}z_{13}z_{23}}\  \Big( \delta^N_{[K} \delta_{L]}^{[P} \delta^{R]M} - \delta^{N[P} \delta^{R]}_{[K} \delta^M_{L]} \Big) \,,
\end{array}
\eeq
where we place indices per convenience, as the internal metric is Euclidean. Moreover, using (\ref{current_4}) and (\ref{sc_split}), we also calculate that
\beq \label{charge_4}
\mathcal{J}^{MN}(z) T_{\textrm{F,int}}(w) \sim \frac{1}{2\sqrt{\alpha'}} \frac{1}{z-w} \Psi^{[M} \partial Z^{N]}(w)
\eeq
and so demonstrate that $T_{\textrm{F,int}}$ again splits into components of disparate charges; the precise form of the split we will not need, namely the OPE (\ref{charge_4}) suffices for the states and amplitudes that we will consider in this work, as we will show.
\end{enumerate}
For the normalisations in (\ref{km_1}), (\ref{km_2}) and (\ref{nonabelian_sc_current}) we follow the conventions of \cite{Feng:2012bb}, while for (\ref{current_4}) and (\ref{km_4}) we have defined the numerical prefactor of the current per convenience. 

In all of the aforementioned cases, (part of)  $\mathfrak{g}$ essentially becomes the $R$--symmetry of the supersymmetry in question, namely $R=\mathfrak{u}(1) $, $R=\mathfrak{su}(2) \times \mathfrak{u}(1)$ and $R=\mathfrak{su}(4)\times \mathfrak{u}(1)$ for $\mathcal{N}=1$, $\mathcal{N}=2$ and $\mathcal{N}=4$ respectively. 
As a final comment we want to mention that states with internal KK momentum break the internal $SO(6)$  $R$--symmetry to $SO(5)$ giving rise to a non--vanishing central charge $Z^{MN}$. In this case the $SO(6)$ current (\ref{current_4}) receives an extra term \cite{Ferrara:1989ud}.

The (on--shell) field content of the open compactified superstring spectrum's first couple of levels has already been worked out in \cite{Feng:2010yx, Feng:2012bb} and we summarise it in terms of supermultiplets as follows:
\begin{itemize}
\item level $0$: it always contains a single massless vector multiplet, whose structure is per case as follows
\begin{subequations}
\begin{align}
\mathcal{N}&=1 : \quad (\textcolor{blue}{1},1/2) \quad , \quad  \textbf{2}_{\textrm{B}}+\textbf{2}_{\textrm{F}} \label{vectorm_1} \\ 
\mathcal{N}&=2  : \quad \big(\textcolor{blue}{1},2(1/2),2(0) \big) \quad , \quad  \textbf{4}_{\textrm{B}}+\textbf{4}_{\textrm{F}}\label{vectorm_2} \\
\mathcal{N}&=4:  \quad \big(\textcolor{blue}{1},4(1/2),6(0) \big) \quad , \quad  \textbf{8}_{\textrm{B}}+\textbf{8}_{\textrm{F}} \label{vectorm_4} \,,
\end{align}
\end{subequations}
\item level $1$: it always contains one massive spin--2 multiplet, whose structure is per case as follows
\begin{subequations}
\begin{align}
\mathcal{N}&=1 : \quad \big(\textcolor{blue}{2},2(3/2),\textcolor{red}{1}\big) \quad , \quad  \textbf{8}_{\textrm{B}}+\textbf{8}_{\textrm{F}} \label{sm_1} \\ 
\mathcal{N}&=2  : \quad \big(\textcolor{blue}{2},4(3/2),6(\textcolor{red}{1}),4(1/2),0\big) \quad , \quad \textbf{24}_{\textrm{B}}+\textbf{24}_{\textrm{F}} \label{sm_2} \\
\mathcal{N}&=4:  \quad \big(\textcolor{blue}{2},8(3/2),27(\textcolor{red}{1}),48(1/2),42(0)\big) \quad , \quad \textbf{128}_{\textrm{B}}+\textbf{128}_{\textrm{F}} \label{sm_4} \,,
\end{align}
\end{subequations}
Moreover, in the $\mathcal{N}=1$ and $\mathcal{N}=2$ cases, there are further two massive chiral multiplets $ \big(1/2,2(0)\big)$,  $\textbf{2}_{\textrm{B}}+\textbf{2}_{\textrm{F}}$ each, and two massive vector multiplets $ \big(1,4(1/2),5(0)\big)$,  $\textbf{8}_{\textrm{B}}+\textbf{8}_{\textrm{F}}$ each, present respectively. 
\end{itemize}
where we have presenting the d.o.f. counting by taking into account the dimensions of the various states under the respective $R$--symmetry.

Interestingly, the massless vector and the massive spin--2, highlighted in blue above and belonging to the \textit{leading} Regge trajectory, take the very \textit{same} form \cite{Feng:2010yx, Feng:2012bb} in all three cases and one that is identical to the structure of their $10D$ ancestors (\ref{gaugevo}) and (\ref{vertop}):
\begin{equation} \label{gaugevo_4d}
V_A^{(-1)}(z,\epsilon,p) = g_{\textrm{o}}\, T^a\, e^{-\phi(z)} \, \epsilon_\mu \psi^\mu(z) \, e^{ipX(z)} \quad , \quad p^2=0 \quad , \quad p \cdot \epsilon = 0
\end{equation}
and
\begin{equation} \label{massive_graviton_open}
\begin{array}{ccl}
V_{B}^{(-1)}(z,B,p)=\displaystyle \frac{g_o}{\sqrt{2\alpha'}}\,\,T^a \, e^{-\phi(z)} \,  B_{\mu \nu} \, i \partial X^\mu(z) \psi^{\nu} (z) \, e^{ipX(z)} \,,
\crbig
p^2=-\frac{1}{\alpha'} \quad , \quad B_{[\mu\nu]}=0 \quad , \quad p^\mu B_{\mu \nu} = 0\quad , \quad B_\mu^\mu=0
\end{array}
\end{equation}
respectively. The reason behind this uniqueness is most obvious in the $(-1)$ ghost picture, where, for example in the case of the massless vector, the weight--$0$ momentum eigenstate forces the presence of an operator of weight--$1/2$ in the vertex operator; $\psi^\mu$ is further the only vector of $SO(D)$ with this weight both before and after the compactification, for any number of spacetime supersymmetries preserved. We may argue that this phenomenon can be observed for all states of the leading Regge trajectory. Note that it is precisely the lowest--lying massive spin--2 state (\ref{massive_graviton_open}) of the leading Regge trajectory whose $3$--point scattering amplitudes we investigated in \cite{Lust:2021jps}.

Turning now to the massive vectors: these are the lowest--lying massive vectors in four dimensions. Since we aim at formulating a double copy that mimics the massless one, which employs the massless vector, we focus on the massive vectors with vertex operators that bear the most resemblance to (\ref{gaugevo_4d}), namely the $U(1)$ singlet of (\ref{sm_1}) \cite{Feng:2010yx} 
\begin{equation} \label{massive_vector_1}
\begin{array}{ccl}
V_{\bm{A}}^{(-1)}(z,a,p)=g_o \, T^a \, e^{-\phi(z)} \,  a_\mu \,\psi^{\mu} (z) \mathcal{J}(z) \, e^{ipX(z)} 
\crbig
p^2=-\frac{1}{\alpha'} \quad , \quad p \cdot a = 0\,
\end{array}
\end{equation}
the $SU(2)$ triplet of (\ref{sm_2}) \cite{Feng:2012bb} 
\begin{equation} \label{massive_vector_2}
\begin{array}{ccl}
V_{\bm{A}}^{(-1)}(z,a,p)=g_o \, T^a \, e^{-\phi(z)} \,  a_\mu^A \,\psi^{\mu} (z) \mathcal{J}^A(z) \, e^{ipX(z)} 
\crbig
p^2=-\frac{1}{\alpha'} \quad , \quad p \cdot a^A = 0
\end{array}
\end{equation}
and the $SU(4)$ vector (i.e. $15$--dimensional representation) of (\ref{sm_4})  \cite{Feng:2012bb} 
\begin{equation} \label{massive_vector_4}
\begin{array}{ccl}
V_{\bm{A}}^{(-1)}(z,a,p)=g_o \, T^a \, e^{-\phi(z)} \,  a_\mu^{MN} \,\psi^{\mu} (z) \mathcal{J}^{MN}(z) \, e^{ipX(z)} 
\crbig
p^2=-\frac{1}{\alpha'} \quad , \quad p \cdot a^{MN} = 0\,.
\end{array}
\end{equation}
Notice that although (\ref{massive_vector_1}), (\ref{massive_vector_2}) and (\ref{massive_vector_4}) belong to the \textit{same} supermultiplet as the massive spin--2 (\ref{massive_graviton_open}) in each case, the massive vectors belong to \textit{subleading} Regge trajectories. Notice also that the vertex operators of these massive vectors are minimal deformations of the massless one: in both cases the spacetime vector $\psi^\mu$ appears, while the momentum eigenstate, having a vanishing conformal weight at the massless level, acquires a weight of $-1$ in the 1st massive level, so that the presence of the respective Kac--Moody current of weight $1$ guarantees that the full vertex operator is physical. We further highlight that, for the $\mathcal{N}=2$ and $\mathcal{N}=4$ cases, there are more massive vectors present at the same level as discussed; these we do not consider, as the structure of their vertex operators deviates schematically significantly from (\ref{gaugevo_4d}).

Finally, it will later be instructive to compare with compactifications of the bosonic string to four dimensions, assuming again (\ref{no_leak}) so that the mass spectrum (\ref{Regge_open_b}) remains unchanged. In this case, the massless vector (level $1$) and the lowest--lying massive vector (level $2$) take, as should be obvious from their ancestors (\ref{gaugevo_open_b}) and (\ref{vertop_open_b}) in $26$ dimensions, the form
\begin{equation} \label{gaugevo_open_b_comp}
V_A (z,\epsilon,p) = \frac{g_{\textrm{o}}}{\sqrt{2\alpha'}} \, T^a\, \epsilon_\mu \partial X^\mu(z) \, e^{ipX(z)}\ \quad , \quad p^2=0 \quad , \quad p \cdot \epsilon = 0\,,
\end{equation}
and
\beq \label{massive_vector_b}
\begin{array}{ccl}
V_{\bm{A}} (z,a,p)= \frac{g_{\textrm{o}}}{\sqrt{2\alpha'}} \, T^a\, a_\mu \,i \partial X^\mu (z)\mathcal{J}_{\textrm{bos}}^M(z) \ e^{ipX(z)} \quad , \quad p^2 = - \frac{1}{\alpha'}  \quad , \quad  p \cdot a =0 \,,
\end{array}
\eeq
where
\beq \label{bosonic_current}
\begin{array}{ccl}
\mathcal{J}_{\textrm{bos}}^M(z)=\frac{1}{\sqrt{2\alpha'}} i\partial Z^M(z)\,,
\end{array}
\eeq
is the $\mathfrak{u}(1)^6$ Kac--Moody current. Using (\ref{bosonic_current}) and the 2--point function (\ref{open_opes_four}) of the internal bosons, we notice that
\beq \label{km_bosonic}
\langle \mathcal{J}^M(z_1)\mathcal{J}^N(z_1) \rangle = \frac{ \delta^{MN}}{z_{12}^2} \quad , \quad \langle \mathcal{J}^M(z_1)\mathcal{J}^N(z_2)\mathcal{J}^K(z_3)  \rangle = 0\,,
\eeq
 the situation is, therefore, very much reminiscent of the case of the $\mathcal{N}=1$ compactification of the superstring. 
 
Importantly, unlike the bosonic (\ref{bosonic_current}) and $\mathcal{N}=4$ (\ref{current_4}) cases, the precise dependence of the $U(1)$ current $\mathcal{J}$ and the $SU(2)$ currents $\mathcal{J}^A$ on the internal fields is irrelevant; knowledge of only their algebra, as well as of their OPE with the respective supercurrent, is necessary, which is why we can think of the massive vectors (\ref{massive_vector_1}) and (\ref{massive_vector_2}) as \textit{universal}. However, in the Abelian orbifold example, which preserves $\mathcal{N}=1$ brane supersymmetry, the $U(1)$ current takes the explicit form \cite{Ferrara:1989ud, Dixon:1989fj}
\beq \label{current1}
\mathcal{J}(z)=  \sum_{i=1}^3\Psi^i \,\ov{\Psi}^{\ov{i}} (z)
\eeq
where the fields  $\Psi^i$ are the complexified internal fermions. Using their 2--point function and the form (\ref{current1}), we can easily show that $\mathcal{J}$ indeed obeys the Abelian algebra (\ref{km_1}). At last, let us emphasise  that there are \textit{two} (commuting) gauge algebras associated with the massive vectors (\ref{massive_vector_1}), (\ref{massive_vector_2}) and (\ref{massive_vector_4}): the Chan--Paton $T^a$ \textit{and} the respective Kac--Moody current referring to the SUSY algebra.

\subsection{Construction of a closed string massive spin--2 state}

As is largely perceived to be the case, closed string states can be constructed by taking the tensor product of two (not necessarily identical) open string states, one to account for left-- and one for right--movers, at the \textit{same} level. This implies that a closed string vertex operator can be written in a factorised form, which is thought to be possible at \textit{any} level, as argued (for the bosonic string) for example in \cite{Green:1987sp}, so that we can write
\beq \label{product_vo}
V^{(q,q)}_{n,\textrm{closed}}(z,\ov{z}, k)  = V^{(q)}_{n,\textrm{open}}(z,p)\, \widetilde{V}^{(q)}_{n,\textrm{open}}(\ov{z},p) \quad ,\quad p^m=\frac{1}{2}k^m \,,
\eeq
where left-- and right--movers are associated with a holomorphic and an antiholomorphic, bearing a tilde, vertex operator respectively. (\ref{product_vo}) implies that gluing two open strings together amounts to taking the trace of the product of the respective $T^A$'s, so that the resulting closed string states are neutral, as expected. To each $V_{n,\textrm{closed}}$ there also corresponds one power of the closed string coupling $g_{\textrm{c}}$, so that $g_{\textrm{c}} \sim g^2_{\textrm{o}}$. 

To avoid confusion, we always denote by $p$ and $k$ open and closed string momenta respectively in what follows, and the corresponding momentum eigenstates with $e^{ipX(z)}$ and $e^{ikX(z, \ov{z})}$. Notice then that (\ref{product_vo}) reproduces the correct conformal dimension $(1,1)$ of physical closed string states according to (\ref{physical}); to this end, it is essential that the two open strings carry \textit{half} the momentum of the resulting closed string, as we illustrate in appendix \ref{conv_app}. Interestingly, this consideration is \textit{redundant} for massless states and is sometimes overlooked in the literature; nevertheless, it is crucial to obtaining physical massive closed string states, namely their correct Regge trajectories after tensoring:
\beq \label{Regge_closed_bos}
M_{n,\,\textrm{closed}}^2 =\begin{cases}
(n-1)\, \frac{4}{\alpha'} \quad , \quad \textrm{for the bosonic string} \\
 n \,  \frac{4}{\alpha'} \quad , \quad \textrm{for the supersring}\,,
\end{cases}
\eeq
where $n\, \in \mathbb{N}_0$. It is also implicit that all contractions pertaining to amplitudes involving only open string states are to be performed using (\ref{boson_two}), while if only closed string states are involved the correlators
\beq \label{closed_opes_ten}
\begin{array}{ccl}
\langle X^m(z_1) X^n(z_2) \rangle_{S_2} = - \frac{ \alpha'}{2} \eta^{m n} \, \ln z_{12} \quad , \quad \langle \widetilde{X}^m(\ov{z}_1) \widetilde{X}^n(\ov{z}_2) \rangle_{S_2} = - \frac{ \alpha'}{2} \eta^{m n} \, \ln \ov{z}_{12} 
 \end{array}
 \eeq 
are to be employed instead, while the prefactors of all other two-- and three--point functions are identical for open and closed strings, for example
\beq
  \langle \psi^m(z_1) \psi^n(z_2)  \rangle = \frac{\eta^{mn}}{z_{12}} \quad , \quad  \langle \widetilde{\psi}^m(\ov{z}_1) \widetilde{\psi}^n(\ov{z}_2)  \rangle = \frac{\eta^{mn}}{\ov{z}_{12}}  \,.
\eeq
Notice that the pairs $X,\psi$ and $\widetilde{X}, \widetilde{\psi}$ can be thought of as \textit{decoupled} copies of each other and mixed correlators as $\langle X^m(z) \widetilde{X}^n(\ov{w}) \rangle$ remain irrelevant for the present work, as we do not compute amplitudes involving both closed and open string external states as in \cite{Lust:2021jps}.

In the bosonic string, it should be obvious that the above considerations hold upon ignoring the ghost pictures in (\ref{product_vo}) and all fermions. The (on--shell) content of the bosonic spectrum's first couple of levels is well--known:
\begin{itemize}
\item level $0$: only one tensor product is possible, namely that of (\ref{tachyon_open_b}). The level thus contains a single scalar tachyon propagating $1$ d.o.f., with vertex operator
\beq
V_{\textrm{t,c}}(z,\ov{z},k) = V_{\textrm{t,o}}(z,k/2)\, \widetilde{V}_{\textrm{t,o}}(\ov{z},k/2)     =g_{\textrm{c}}\,   e^{ikX(z,\ov{z})} \quad , \quad k^2=\frac{4}{\alpha'}\,.
\eeq

\item level $1$: only one tensor product is possible, mamely that of (\ref{gaugevo_open_b})
\beq \label{splitd_1}
\bf{24} \otimes \bf{24} = \bf{299} \, \oplus  \, \bf{276} \, \oplus \, \bf{1}.
\eeq
 This level thus contains three states, the vertex operator operator of all of which takes the form
\begin{equation} \label{massless_dc}
\begin{array}{ccl}
V_G (z,\ov{z},\varepsilon,k) &=& V_A(z,\epsilon,k/2) \, \widetilde{V}_A(\ov{z},\widetilde{\epsilon},k/2)
\crbig
&=& \frac{g_{\textrm{c}}}{2\alpha'} \, \varepsilon_{mn} \,  \partial X^m(z,\ov{z}) \ov{\partial} X^n(z,\ov{z}) \, e^{ikX(z,\ov{z})}\ 
\end{array}
\end{equation}
with
\begin{equation} 
\begin{array}{ccl}
 k^2=0 \quad , \quad k^m  \varepsilon_{mn} = 0\,,
\end{array}
\end{equation}
where $\varepsilon_{mn}$ is a transverse tensor of $SO(1,25)$ that splits into the polarisations of the graviton, the Kalb--Ramond and dilaton according to (\ref{splitd_1}).
\end{itemize} 
Now let us consider level $2$. We observe that again only one tensor product is possible, namely that of (\ref{vertop_open_b})
\beq \label{splitd_2}
\bf{324} \otimes \bf{324} = \bf{20150} \, \oplus  \, \bf{32175} \, \oplus \, \bf{52026} \, \oplus \, \textcolor{red}{\bf{324}} \, \oplus \, \bf{300} \, \oplus \, \bf{1} \,,
\eeq
so that this level contains several states (symmetric, antisymmetric and of mixed--symmetry) with vertex operator
\beq \label{massive_other}
\begin{array}{ccl}
V_M (z,\ov{z},\alpha,k) &=& V_B (z,B,k/2) \, \widetilde{V}_B (\ov{z},\widetilde{B},k/2)
\crbig
&=&\frac{ g_{\textrm{c}} }{(2\alpha')^2} \, \alpha_{mnkl} \, \partial X^m (z,\ov{z}) \partial X^n (z,\ov{z}) \ov{\partial} X^k(z,\ov{z})\ov{\partial} X^l(z,\ov{z})   \, e^{ikX(z,\ov{z})}\,,
\end{array}
\eeq
with
\beq \label{os_cond}
\begin{array}{ccl} 
 k^2=-\frac{4}{\alpha'} \quad , \quad k^m \alpha_{mnkl} =0 \,,
\end{array}
\eeq
where $\alpha_{mnkl}$ is a transverse tensor of $SO(1,25)$ that splits according to (\ref{splitd_2}); in particular, its symmetric traceless component, highlighted in red, is the \textit{lowest} lying massive ``graviton'' of the closed bosonic string spectrum in $26$ dimensions. Notice that it appears in multiplicity equal to $1$.

Turning to the closed superstring spectrum: it is  organized in multiplets of (spacetime) $\mathcal{N}=2$, $D=10$ supersymmetry. Focusing on the NS--NS sector, the (on--shell) content of the spectrum's level $0$ is a well--known example of (\ref{product_vo}) and consists in one possible tensor product, namely that of two massless vectors (\ref{gaugevo}):
\beq \label{splitd_s1}
\bf{8} \otimes \bf{8} = \bf{35} \oplus \bf{28} \oplus \bf{1}\,.
\eeq
It contains three states, the vertex operator of all of which takes the form
\beq \label{graviton_vo_ten}
\begin{array}{ccl}
V_G^{(-1,-1)}(z,\ov{z},\varepsilon,k) &=& V_A^{(-1)}(z,\epsilon,k/2) \, \widetilde{V}_A^{(-1)}(\ov{z},\widetilde{\epsilon},k/2)
\\
&=& g_{\textrm{c}} \, \varepsilon_{mn} \, e^{-\phi(z)-\widetilde{\phi}(\ov{z})} \,  \psi^m(z) \widetilde{\psi}^n(\ov{z}) \, e^{ikX(z, \ov{z})}\,,
\end{array}
\eeq
with
\beq
 k^2 = 0 \quad , \quad k^m \varepsilon_{mn} =0 \,,
\eeq
where $ \varepsilon_{mn}$ is a transverse tensor of $SO(1,9)$ that splits according to (\ref{splitd_s1}) into the polarisations of the graviton, Kalb--Ramond and dilaton. 

Now let us consider level $1$. Using (\ref{vertop1}) and (\ref{vertop}), we observe that there are three tensor products possible, namely
\begin{subequations}
\begin{align}
\bf{84} \otimes \bf{84} &= \bf{2772} \, \oplus \, \bf{1980} \, \oplus \, \bf{924} \, \oplus \, \bf{594} \, \oplus \, \bf{495} \, \oplus \bf{126} \, \oplus \, \bf{84} \, \oplus \, \textcolor{red}{\bf{44}} \, \oplus \, \bf{36} \, \oplus \, \bf{1} \\
\bf{44} \otimes \bf{84} &=\bf{2457} \, \oplus \, \bf{924} \, \oplus \, \bf{231} \, \oplus \, \bf{84} \\
    \bf{44} \otimes \bf{44} &= \bf{910} \, \oplus \, \bf{495} \, \oplus \, \bf{450} \, \oplus \, \textcolor{red}{\bf{44}} \, \oplus \, \bf{36} \, \oplus \, \bf{1}\,, \label{third_sp}
\end{align}
\end{subequations}
that correspond to 
\begin{subequations}
\begin{align}
V_I^{(-1,-1)}(z,\ov{z},k) &\equiv V_E^{(-1)}(z,E,k/2) \, \widetilde{V}_E^{(-1)}(\ov{z},\widetilde{E},k/2) \\
V_{II}^{(-1,-1)}(z,\ov{z},k) &\equiv V_B^{(-1)}(z,B,k/2) \, \widetilde{V}_E^{(-1)}(\ov{z},\widetilde{E},k/2) \\
V_{III}^{(-1,-1)}(z,\ov{z},k) &\equiv V_B^{(-1)}(z,B,k/2) \, \widetilde{V}_B^{(-1)}(\ov{z},\widetilde{B},k/2) \label{third_vo} \,,
\end{align}
\end{subequations}
respectively. Notice that the states highlighted above in red are the lowest--lying massive ``gravitons'' of the closed superstring; the massive spin--2 state appears at this level in multiplicity equal to $2$. For example, using (\ref{third_vo}), the vertex operator of massive spin--2 state of (\ref{third_sp}) takes the form
\beq
\begin{array}{ccl}
V_M^{(-1,-1)} (z,\ov{z},\alpha,k)&=& \frac{g_{\textrm{c}}}{2\alpha'} \, \alpha_{mnkl} \, e^{-\phi(z)-\widetilde{\phi}(\ov{z})} 
\crbig
&& \qquad \times  \partial X^m (z,\ov{z}) \psi^n(z)  \, \ov{\partial} X^k(z,\ov{z})  \widetilde{\psi}^l (\ov{z}) \, e^{ikX(z,\ov{z})}
\end{array}
\eeq
with
\beq
\begin{array}{ccl}
 k^2=-\frac{4}{\alpha'} \quad , \quad k^m \alpha_{mnkl} =0 \,,
\end{array}
\eeq
where $\alpha_{mnkl}$ has to be suitably constrained to account for the $\bf{44}$ representation of $SO(9).$ 

Turning to compactifications to four dimensions, let us note first that the assumption (\ref{no_leak}) implies
\beq \label{no_leak_2}
k^m =(k^\mu,0)\,.
\eeq
Other than that, the rule (\ref{product_vo}) remains valid, so that the closed string spectrum can be constructed as before compactifications, using the tools of section \ref{section_comp_open} supplemented by their antiholomorphic counterparts. For all cases of supersymmetry preserved, namely $\mathcal{N}=2,4,8$ in the bulk by means of the respective brane products
\begin{subequations}
\begin{align}
\mathcal{N}&=1\otimes\mathcal{N}=1 \label{tp_1} \\
\mathcal{N}&=2\otimes\mathcal{N}=2 \label{tp_2} \\
\mathcal{N}&=4\otimes\mathcal{N}=4 \label{tp_4}\,,
\end{align}
\end{subequations}
 level $0$ consists in a single possible tensor product, that of (\ref{gaugevo_4d}) that is well--known
\beq \label{graviton_vo}
\begin{array}{ccl}
V_G^{(-1,-1)}(z,\ov{z},\varepsilon,k) &=& V_A^{(-1)}(z,\epsilon,k/2) \, \widetilde{V}_A^{(-1)}(\ov{z},\widetilde{\epsilon},k/2)
\\
&=&g_{\textrm{c}} \, \varepsilon_{\mu \nu} \, e^{-\phi(z)-\widetilde{\phi}(\ov{z})} \,  \psi^\mu(z) \widetilde{\psi}^\nu(\ov{z}) \, e^{ikX(z, \ov{z})}\,,
\end{array}
\eeq
The polarisation $\varepsilon_{\mu \nu}$ is a (constrained) tensor of $SO(2)$ and splits into the irreducible representations that correpond to the graviton, Kalb--Ramond and dilaton in four space--time  dimensions \cite{Mayr:1993vu}. We find it instructive to write that of the graviton in terms of the massless polarisations $\epsilon_\mu$ and $\widetilde{\epsilon}_\mu$  
 \beq \label{pol_G_prod}
\begin{array}{ccl}
 \varepsilon_{\mu\nu} &=& \frac{1}{2}\big(\epsilon_\mu \widetilde{\epsilon}_\nu + \epsilon_\nu \widetilde{\epsilon}_\mu \big) - \frac{1}{8} \epsilon \cdot \widetilde{\epsilon} \, \big( \eta_{\mu\nu} - k_\mu \zeta_\nu - k_\nu \zeta_\mu \big)\,,
 \end{array}
\eeq
where $\zeta^\mu$ is an auxiliary vector that satisfies $k \cdot \zeta = 1$ and we have imposed $ \varepsilon_{\mu\nu} = \varepsilon_{\nu\mu} \,,\, \varepsilon_\mu^\mu =0 \,$. Notice that while these two conditions are built in the respective irreducible represenation of $SO(2)$, the rest of the graviton's properties are \textit{inherited} from the vector:
\begin{enumerate}
\item masslessness and transversality: 
\beq \label{cond_on1}
p^2 =0 \, \Rightarrow \, k^2 = 0 \quad , \quad p \cdot \epsilon = p \cdot{\widetilde{\epsilon}}=0 \, \Rightarrow  \, k^\mu \varepsilon_{\mu\nu} = k^\nu \varepsilon_{\mu\nu}=0\,.
\eeq
\item linarized diffeos from \textit{simultaneous}  gauge transformations:
\beq
\epsilon_\mu \rightarrow \epsilon_\nu +p_\mu \quad \textrm{and}\quad \widetilde{\epsilon}_\mu \rightarrow \widetilde{\epsilon}_\mu +p_\mu \quad \Rightarrow \quad \varepsilon_{\mu\nu} \rightarrow \varepsilon_{\mu\nu} + k_\mu \xi_\nu + k_\nu \xi_\mu
\eeq
where
\beq
\xi_\mu \equiv \frac{1}{4}\, (\epsilon_\mu +\widetilde{\epsilon}_\mu) \quad, \quad p \cdot \epsilon = p \cdot{\widetilde{\epsilon}}=0 \, \Rightarrow  \, k \cdot \xi =0\,.
\eeq
\end{enumerate}

Proceeding to level $1$, there are now various possibilities depending on the amount of supersymmetry preserved. For example, for $\mathcal{N}=2$ in the bulk:
\begin{subequations}
\begin{align}
\bf{3}\otimes \bf{3}&=\textcolor{red}{\bf{5}}\oplus\bf{3}\oplus\bf{1} \label{tp_1} \\
\bf{5}\otimes \bf{3}&=\bf{7}\oplus\textcolor{red}{\bf{5}}\oplus\bf{3} \label{tp_2} \\ 
\bf{5} \otimes \bf{5}&=\bf{9}\oplus\bf{7}\oplus\textcolor{red}{\bf{5}}\oplus\bf{3}\oplus\bf{1}\,, \label{tp_3}
\end{align}
\end{subequations}
so that the massive spin--2 in this case appears in multiplicity $3$. As explained in the previous subsection, we will focus on tensor products of vectors as (\ref{tp_1}), so that the vertex operator of the massive spin--2 we generically construct as
\beq \label{dc_massive}
\begin{array}{ccl}
V_M^{(-1,-1)}(z,\ov{z},\alpha,k) &=& V_{\bm{A}}^{(-1)}(z,a,k/2) \, \widetilde{V}_{\bm{A}}^{(-1)}(\ov{z},\widetilde{a},k/2)\,.
\end{array}
\eeq
Using (\ref{massive_vector_1}), (\ref{massive_vector_2}) and (\ref{massive_vector_4}), we then have for bulk $\mathcal{N}=2,4,8$ respectively
\begin{numcases}{V_M^{(-1,-1)}(z,\ov{z},\alpha,k) =  g_{\textrm{c}}  \,  e^{-\phi(z)-\widetilde{\phi}(\ov{z})} \,\psi^\mu(z) \, \widetilde{\psi}^\nu(\ov{z}) \, e^{ikX(z, \ov{z})} }
\alpha_{\mu \nu}\, J(z) \,  \widetilde{J}(\ov{z})\label{dc_massive_1} \\
\alpha_{\mu \nu}^{AA'}\, J^A(z) \,  \widetilde{J}^{A'}(\ov{z})\label{dc_massive_2} \\
\alpha_{\mu \nu}^{MNM'N'} \, J^{MN}(z) \,  \widetilde{J}^{M'N'}(\ov{z}) \label{dc_massive_4}\,.
\end{numcases}
Notice that the on--shell mass condition and transversality are again \textit{inherited} from the vector:
\beq \label{cond_on3}
p^2 =-\frac{1}{\alpha'} \, \Rightarrow \, k^2 = -\frac{4}{\alpha'} \quad , \quad p \cdot a = p \cdot{\widetilde{a}}=0 \, \Rightarrow  \, k^\mu \alpha_{\mu \nu} = k^\nu \alpha_{\mu \nu}=0\,,
\eeq
which holds for any of the gauge currents. Let us further highlight that the worldsheet structure of (\ref{dc_massive_1})--(\ref{dc_massive_4}) \textit{mimics} that of the massless graviton's vertex operator (\ref{graviton_vo}): our massive spin--2 states are the lowest--lying massive states that resemble the graviton the most. Interestingly, they all belong to \textit{subleading} Regge trajectories.

\begin{table} 
\centering 
\renewcommand{\arraystretch}{1.5}
  \begin{tabular}{ c || c   }
   little group & $\ydiagram{1} \otimes \ydiagram{1} = \ydiagram{2} \oplus \ydiagram{1,1} \oplus \bullet$    \\ \hline
   $SO(2)$ & $2 \otimes 2 = 2 \oplus 1 \oplus 1 $  \\ \hline
   $SO(3)$ & $3 \otimes 3 = 5 \oplus 3 \oplus 1$ 
  \end{tabular}
\renewcommand{\arraystretch}{1}
\caption{Tensoring of massless and massive vectors}\label{tens}
\end{table}
For the $\mathcal{N}=2$ bulk case (\ref{tp_1}), we illustrate the similarities of the tensor decomposition of the products of massless and of massive vectors in table \ref{tens}. For the massive spin--2 state (\ref{dc_massive_1}) that employs the $U(1)$ current, we impose that $\alpha_{\mu \nu}$ be the symmetric traceless representation of $SO(3)$ that further respects (\ref{cond_on3}) and find that
\beq \label{massive_pol}
\begin{array}{ccl}
 \alpha_{\mu \nu} &=& \frac{1}{2}\big(a_\mu \widetilde{a}_\nu + a_\nu \widetilde{a}_\mu \big) - \frac{1}{3} a \cdot \widetilde{a} \, \big( \eta_{\mu \nu} +\frac{k_\mu k_\nu}{m^2}\big)\,.
 \end{array}
\eeq
At this point, it is instructive to compare with the massive double copy of \cite{Momeni:2020vvr, Johnson:2020pny, Engelbrecht:2022aao}. Crucially, we observe that (\ref{massive_pol}) matches the helicity decomposition of the massive spin--2 state constructed in \cite{Momeni:2020vvr, Johnson:2020pny} by means of \textit{Yang--Mills fields} with Proca--like mass terms, if $a_\mu$ and $\widetilde{a}_\mu$ are taken to be their respective polarisations.

At this point, let us stress that the massive spin--$2$ states we have constructed, which fall into supermultiplets of $\mathcal{N}=2,4,8$, $D=4$ respectively, are charged under the respective Kac--Moody internal currents, as is manifest in their vertex operators (\ref{dc_massive_1})--(\ref{dc_massive_4}). Consequently, they are \textit{not} singlets under the respective $R$--symmetry, that is  $SU(2)$, $USp(8)$ and $USp(16)$ \cite{Ferrara:2018iko} correspondingly, and their supermultiplets further contain \textit{higher--spin} states, which are the ones that transform as singlets. It is beyond the scope of this work to construct these supermultiplets and determine the precise $R$--symmetry representations of their components, but we would like to illustrate this setup in the $\mathcal{N}=2$ example. In particular,  (\ref{tp_1})--(\ref{tp_3}) contain three massive spin--2 bulk states, with those of (\ref{tp_1}) and (\ref{tp_2}) being associated with the current $\mathcal{J}$. Consequently, these spin--2 states of (\ref{tp_1}) and (\ref{tp_2}) are not the highest--spin components in their supermultiplets and the latter should further contain the higher--spin states of (\ref{tp_3}) that propagate $7$ and $9$ d.o.f. respectively and which are singlets by construction. Instead, the spin--2 state of (\ref{tp_3}), which belongs to a \textit{leading} Regge trajectory, has vertex operator 
\beq \label{leading_r}
\begin{array}{ccl}
V_{\widetilde{M}}^{(-1,-1)} (z,\ov{z},\alpha,k) &=&  \frac{g_{\textrm{c}}}{2\alpha'} \, \alpha_{\mu \nu \kappa \lambda} \,  e^{-\phi(z)-\widetilde{\phi}(\ov{z})}
\crbig
&& \qquad  \times  \, i\partial X^\mu(z,\ov{z}) \psi^\nu(z) \, i\ov{\partial} X^\kappa(z,\ov{z}) \widetilde{\psi}^\lambda(\ov{z}) \, e^{ikX(z, \ov{z})} \,,
\end{array}
\eeq
with a suitably constrained $\alpha_{\mu \nu \kappa \lambda}$ to account for the d.o.f. of a symmetric rank--2 tensor; (\ref{leading_r}) is manifestly a singlet under $SU(2)$ and so the highest--spin component of a massive supermultiplet of $\mathcal{N}=2$. Multiplets for massive spin--2 states of extended supersymmetry that are $R$--symmetry singlets have been constructed within both string \cite{Ferrara:2018iko} and field theory \cite{Zinoviev:2002xn}. Moreover, it is precisely as singlets that massive spin--2 states are built in a supersymmetrizable massive double copy to cubic order in field theory \cite{Engelbrecht:2022aao}. However, in this work, we focus on massive spin--2 states with a (double copy) structure that resembles most that of the graviton, namely the vertex operators (\ref{dc_massive_1})--(\ref{dc_massive_4}). Our states are thus inevitably \textit{charged} under $R$--symmetry.

The corresponding space--time bulk SUSY 
charges referring to (\ref{tp_1})--(\ref{tp_4}) can be constructed from the fields (\ref{split1}) and their anti--holomorphic counterparts by some contour integrals in the complex plane. Likewise the resulting half number of SUSY charges
 on the Dp--brane world--volume can be represented by the single holomorphic sector  (\ref{split1})
 subject to the doubling trick on the disk \cite{Stieberger:2007jv}.

We are almost ready to proceed to computing scattering amplitudes involving the massive spin--2 states (\ref{dc_massive_1})--(\ref{dc_massive_4}) we have constructed; however, we first need these vertex operators in different pictures. In particular, for any open string amplitude, the total ghost picture has to be equal to $(-2)$ to cancel the ghost charge of the disk background, while for closed string amplitudes to total ghost picture has to be equal to $(-2,-2)$. It suffices to apply the picture changing operation on open string states, so we begin with vectors and, more specifically, we will first compute $V_{\bm{A}}^{(-1)}(z,a,p)$ for $\mathcal{N}=1,2,4$; the massive spin--2 vertex operators $V_M^{(0,0)}(z,\ov{z},\alpha,k)$ will then be given by the respective tensor products. Generically, picture changing is effectuated via \cite{Friedan:1985ey}
\beq
V^{(q+1)}(w) = \lim_{z \rightarrow w} P_F(z) \, V^{(q)}(w)\,,
\eeq
where
\beq
P_F(z) \equiv e^{\phi(z)} T_F(z)
\eeq
is the picture changing operator that is essentially a (worldsheet) supersymmetry transformation and we have ignored contributions from the rest of the terms in the BRST charge (\ref{brst_compo}), as they have no effect on the states that we consider in the present work.

After the compactification, $T_F(z)$ is given by (\ref{sc_split}). Consequently, using $V_{A}^{(-1)}(w,a,p)$ of (\ref{gaugevo_4d}), we have that
\beq
V_{A}^{(0)}(w,\epsilon,p)= g_{\textrm{o}} \, T^a\, \epsilon_\mu \, \lim_{z \rightarrow w} e^{\phi(z)}  \bigg[ \frac{i}{2\sqrt{2\alpha'}}\, \psi \partial X \, (z)+ T_{\textrm{F,int}}(z) \bigg] e^{-\phi(w)} \,  \psi^\mu(w) \, e^{ipX(w)} \, 
\eeq
so for all cases $\mathcal{N}=1,2,4$, since $T_{\textrm{F,int}}(z) $ is decoupled from matter fields,
\beq \label{massless_vector_pc}
V_{A}^{(0)}(z,\epsilon,p) = \frac{g_{\textrm{o}}}{\sqrt{2\alpha'}} \, T^a \,\epsilon_\mu \big[i\partial X^\mu(z) +2\alpha' \, (p \cdot \psi) \psi^\mu(z) \big] e^{ipX(z)}\,,
\eeq
where we have used the OPEs of appendix \ref{app_ope}. Turning to the massive vectors, we start with the $\mathcal{N}=1$ case (\ref{massive_vector_1}): using (\ref{split_1}) we have that 
\beq
\begin{array}{ccl}
V_{\bm A}^{(0)}(w,a,p)= g_{\textrm{o}} \, T^a\, a_\mu \, \displaystyle \lim_{z \rightarrow w} e^{\phi(z)}  \bigg\{ \frac{i}{2\sqrt{2\alpha'}}\, \psi \partial X \, (z)+ \frac{1}{\sqrt{2}}\big[T_{\textrm{F,int}}^+(z) + T_{\textrm{F,int}}^-(z) \big] \bigg\}
\crbig
\qquad \qquad   \times \, e^{-\phi(w)} \,  \psi^\mu(w) \mathcal{J}(w) \, e^{ipX(w)} \, 
\end{array}
\eeq
from which, using the the OPE (\ref{ope_current_superc}) and those of appendix \ref{app_ope}, we find
\beq \label{massive_vector_pc_1}
\begin{array}{ccl}
V_{\bm{A}}^{(0)}(z,a,p) = g_{\textrm{o}} \,  T^a a_\mu \, \bigg\{  \frac{i}{2\sqrt{2\alpha'}} \big[i\partial X^\mu(z) +2\alpha' \, (p \cdot \psi) \psi^\mu(z) \big] \, \mathcal{J}(z)
\crbig
\qquad \qquad \qquad  + \frac{1}{\sqrt{6}} \, \psi^\mu \,  \big[T_{\textrm{F,int}}^+ - T_{\textrm{F,int}}^- \big]  \bigg\} \, e^{ipX(z)} \,.
\end{array}
\eeq

We continue with the $\mathcal{N}=2$ case (\ref{massive_vector_2}): using (\ref{split_2}) and (\ref{sc_2_doublet}) we have that
\beq 
\begin{array}{ccl}
V_{\bm A}^{(0)}(w,a,p)= g_{\textrm{o}} \, T^a\, a_\mu^A \, \displaystyle \lim_{z \rightarrow w} e^{\phi(z)}  \bigg\{ \frac{i}{2\sqrt{2\alpha'}}\, \psi \partial X \, (z)+ T_{\textrm{F,int}}^{c=3}(z) +  \frac{1}{\sqrt{2}} \, \lambda^i(z) \, g_i(z)  \bigg\}
\crbig
\qquad \qquad   \times \, e^{-\phi(w)} \,  \psi^\mu(w) \mathcal{J}^A(w) \, e^{ipX(w)} \, 
\end{array}
\eeq
so, employing the OPE (\ref{nonabelian_sc_current}) and those of appendix \ref{app_ope}, and noticing that $T_{\textrm{F,int}}^{c=3}$ and $g_i$ are \textit{decoupled} from the rest of the fields present in $V_{\bm A}^{(-1)}(w,a,p)$, we find that
\beq \label{massive_vector_pc_2}
\begin{array}{ccl}
V_{\bm{A}}^{(0)}(z,a,p) = g_{\textrm{o}} \,  T^a a^A_\mu \, \bigg\{  \frac{i}{\sqrt{2\alpha'}} \big[i\partial X^\mu(z) +2\alpha' \, (p \cdot \psi) \psi^\mu(z) \big] \, \mathcal{J}^A(z)
\crbig
\qquad \qquad \qquad  +\psi^\mu \, g_i \,  (\tau^A)^i_j \, \lambda^j  \bigg\} \, e^{ipX(z)} \,.
\end{array}
\eeq
Finally, for the $\mathcal{N}=4$ case (\ref{massive_vector_4}): using (\ref{split_2}) and (\ref{sc_2_doublet}) we have that
\beq 
\begin{array}{ccl}
V_{\bm A}^{(0)}(w,a,p)= g_{\textrm{o}} \, T^a\, a_\mu^{MN} \, \displaystyle \lim_{z \rightarrow w} e^{\phi(z)}  \bigg[ \frac{i}{2\sqrt{2\alpha'}}\, \psi \partial X \, (z)+ T_{\textrm{F,int}}(z)   \bigg] 
\crbig
\qquad \qquad \qquad \qquad \qquad \times \, e^{-\phi(w)} \,  \psi^\mu(w) \mathcal{J}^{MN}(w) \, e^{ipX(w)} \, 
\end{array}
\eeq
so that, using the OPE (\ref{charge_4}) and again those of appendix \ref{app_ope}, we find
\beq \label{massive_vector_pc_4}
\begin{array}{ccl}
V_{\bm{A}}^{(0)}(z,a,p) = g_{\textrm{o}} \,  T^a a^{MN}_\mu \, \bigg\{  \frac{i}{\sqrt{2 \alpha'}} \big[i\partial X^\mu(z) +2\alpha' \, (p \cdot \psi) \psi^\mu(z) \big] \, \mathcal{J}^{MN}(z)
\crbig
\qquad \qquad \qquad  +\frac{1}{\sqrt{\alpha'}} \, \psi^\mu \, \Psi^{[M} \partial Z^{N]}  \bigg\} \, e^{ipX(z)} \,.
\end{array}
\eeq
It is then straightforward to obtain the massless and massive spin--2 vertex operators in the $(0,0)$ ghost picture for $\mathcal{N}=2,4,8$ via
\beq \label{dc_zero}
\begin{array}{ccl}
V_G^{(0,0)}(z,\ov{z},\varepsilon,k) &=& V_{A}^{(0)}(z,\epsilon,k/2) \, \widetilde{V}_{A}^{(0)}(\ov{z},\widetilde{\epsilon},k/2)\,.
\crbig
V_M^{(0,0)}(z,\ov{z},\alpha,k) &=& V_{\bm{A}}^{(0)}(z,a,k/2) \, \widetilde{V}_{\bm{A}}^{(0)}(\ov{z},\widetilde{a},k/2)\,.
\end{array}
\eeq
It should be evident that the respective on--shell conditions are the same for any picture of the same physical state.

\section{Three--point amplitudes} \label{amplitudes}

In a string scattering amplitude with $N_o$ open string and $N_c$ closed string external states, there appears the so--called Koba--Nielsen factor, see for example \cite{Stieberger:2009hq}, that is essentially the correlator of the momentum eigenstates of the external states. In particular, it takes the form
\beq \label{KB_open}
\begin{array}{ccl}
\mathcal{E}_{N_o} \equiv \langle \displaystyle  \prod_{i=1}^{N_o} e^{ip_iX(x_i)}  \rangle &=& \displaystyle \prod_{i<j}^{N_o} (x_i-x_j)^{2\alpha' \, p_i \cdot p_j} 
\end{array}
\eeq
for open string amplitudes and 
\beq \label{KB_closed}
\begin{array}{ccl}
E_{N_c} \equiv \langle \displaystyle \prod_{j=1}^{N_c}  e^{ik_j X(x_j,\ov{x}_j)} \rangle &=& \displaystyle  \prod_{i<j}^{N_c} |z_i-z_j|^{\alpha' \, k_i \cdot k_j}
\end{array}
\eeq
for closed string amplitudes. For closed string external states of momentum $k^\mu$, constructed out of pairs of open string states carrying momentum $p^\mu=\frac{1}{2}k^\mu$ according to the rule (\ref{product_vo}), we have that $N_c=2N_o$ and so we observe that
\beq \label{KB_split}
\begin{array}{ccl}
 \displaystyle  \prod_{i<j}^{N_c} |z_i-z_j|^{\alpha' \, k_i \cdot k_j}=  \displaystyle  \prod_{i<j}^{N_c/2} (z_i-z_j)^{\frac{1}{2}\alpha' \, k_i \cdot k_j} \, \prod_{l<r}^{N_c/2}  (\ov{z}_l-\ov{z}_r)^{\frac{1}{2}\alpha' \, k_l \cdot k_r}\,,
\end{array}
\eeq
where the equality holds strictly \textit{because} of the level--matching condition. Using (\ref{KB_open}), (\ref{KB_closed}) and (\ref{KB_split}), we conclude that
\beq \label{KB_Cardy}
E_{N_c} =\mathcal{E}_{N_o} \widetilde{\mathcal{E}}_{N_o}\,,
\eeq
where we have defined
\beq \label{KB_open2}
\begin{array}{ccl}
\widetilde{\mathcal{E}}_{N_o} \equiv \langle \displaystyle  \prod_{i=1}^{N_o} e^{ip_i\widetilde{X}(x_i)}  \rangle &=& \displaystyle \prod_{i<j}^{N_o} (\ov{z}_i-\ov{z}_j)^{2\alpha' \, p_i \cdot p_j} 
\end{array}
\eeq
We recognize (\ref{KB_Cardy}) as essentially a manifestation of Cardy's trick; crucially, it remains valid for both massless and massive string states, of \textit{any} mass level.

We are ready to calculate $3$--point amplitudes $\mathcal{A}_3$ with massless and/or massive vectors and $\mathcal{M}_3$ with massless and/or massive spin--2 states as external states. For each external leg we consider both possible cases, namely four open and four closed string distinct amplitudes. With $p_i$ and $k_i$, $i=1,2,3$ we denote the external momenta in the former and in the latter case respectively, so that momentum conservation takes the form
\beq \label{mom_cons}
\sum_i p_i^\mu = \sum_i k_i^\mu =0\,,
\eeq
for all cases. The ghost pictures of the external states are freely chosen so long as $\mathcal{A}_3$ and $\mathcal{M}_3$ have total picture $(-2)$ and $(-2,-2)$ respectively and we denote with lower indices the worldsheet location of the insertion of the respective operator, for example $c_1=c(z_1)$. To treat the divergences of the amplitudes due to the worldsheet conformal invariance, that is essentially $PSL(2,\mathbb{R})$ and $SL(2,\mathbb{C})$ for open and closed strings respectively, one inserts for all open string vertex operators a $c$--ghost at the same worldsheet location and the same, together with its antiholomorphic counterpart, for all closed string vertex operators. We will also normalise all $\mathcal{A}_3$ and $\mathcal{M}_3$ by means of
\beq
C_{D_2}=\frac{1}{\alpha' g_o^2} \quad , \quad C_{S_2}=\frac{8\pi}{\alpha' g_{\textrm{c}}^2}
\eeq
respectively, as unitarity or equivalently the factorisation of $4$--point amplitudes in terms of the $3$--point amplitudes enforces; see for the tachyon example \cite{Polchinski:1998rr}. In the following, we omit the integrations over the worldsheet locations of the external states already in the beginning, since the worldsheet dependence of $3$--point amplitudes drops out after applying Wick's theorem and using the relevant $2$--point functions, before performing the integrals.

\subsection{Vector scattering}

In this subsection, each external leg is associated with the vertex operator (\ref{gaugevo_4d}) or (\ref{massless_vector_pc}) if massless and with (\ref{massive_vector_1}) or (\ref{massive_vector_pc_1}), (\ref{massive_vector_2}) or (\ref{massive_vector_pc_2}) and  (\ref{massive_vector_4}) or (\ref{massive_vector_pc_4}) if massive for $\mathcal{N}=1,2,4$ respectively. We can thus write schematically
\begin{itemize}
\item three massless vectors: to the best of our knowledge, this is the only case that has been studied before, see for example \cite{Green:1987sp} or \cite{Polchinski:1998rr}
\beq \label{A3_massless}
\begin{array}{ccl}
\mathcal{A}_{AAA}&=&  \langle c_1V_{A}^{(-1)}(z_1,\epsilon_1,p_1) \, c_2V_{A}^{(-1)}(z_2,\epsilon_2,p_2) \, c_3V_{A}^{(0)}(z_3,\epsilon_3,p_3) \rangle
\crbig
&&+ (a,p_1,\epsilon_1) \leftrightarrow (b,p_2,\epsilon_2)
\end{array}
\eeq
with on--shell conditions
\beq
\begin{array}{ccl}
p_i^2 =0 \quad , \quad p_i \cdot \epsilon_i=0 \quad, \quad p_i \cdot p_j =0 \,
\end{array}
\eeq
\item two massless and one massive vector:
\beq \label{A3_massless_massive1}
\begin{array}{ccl}
\mathcal{A}_{AA\bm{A}}  &=& \langle c_1V_{A}^{(-1)}(z_1,\epsilon_1,p_1) \, c_2V_{A}^{(-1)}(z_2,\epsilon_2,p_2) \, c_3V_{\bm{A}}^{(0)}(z_3,a_3,p_3) \rangle
\crbig
&&+ (a,p_1,\epsilon_1) \leftrightarrow (b,p_2,\epsilon_2)
\end{array}
\eeq
with on--shell conditions
\beq
\begin{array}{ccl}
p_1 \cdot \epsilon_1= p_2 \cdot \epsilon_2 = p_3 \cdot a_3=0
\crbig
p_1^2=p_2^2=0 \quad , \quad p_3^2 =-\frac{1}{\alpha'} \quad , \quad p_1 \cdot p_2 = -\frac{1}{2\alpha'} \quad , \quad p_1 \cdot p_3 = p_2 \cdot p_3=\frac{1}{2\alpha'}\,,
\end{array}
\eeq
\item two massive and one massless vector:
\beq \label{A3_massles_massive2}
\begin{array}{ccl}
\mathcal{A}_{\bm{A} \bm{A}A}  &=& \langle c_1V_{\bm{A}}^{(-1)}(z_1,a_1,p_1) \, c_2V_{\bm{A}}^{(-1)}(z_2,a_2,p_2) \, c_3V_{A}^{(0)}(z_3,\epsilon_3,p_3) \rangle
\crbig
&&+ (a,p_1,\epsilon_1) \leftrightarrow (b,p_2,\epsilon_2)
\end{array}
\eeq
with on--shell conditions
\beq 
\begin{array}{ccl}
p_1 \cdot a_1= p_2 \cdot a_2 = p_3 \cdot \epsilon_3=0
\crbig
p_1^2=p_2^2 =-\frac{1}{\alpha'}  \quad , \quad p_3^2=0 \quad , \quad p_1 \cdot p_2 = \frac{1}{\alpha'} \quad , \quad p_1 \cdot p_3 = p_2 \cdot p_3=0\,,
\end{array}
\eeq
\item three massive vectors:
\beq \label{A3_massive}
\begin{array}{ccl}
\mathcal{A}_{\bm{A} \bm{A}\bm{A}} & =&  \langle c_1V_{\bm{A}}^{(-1)}(z_1,a_1,p_1) \, c_2V_{\bm{A}}^{(-1)}(z_2,a_2,p_2) \, c_3V_{\bm{A}}^{(0)}(z_3,a_3,p_3) \rangle
\crbig
&&+ (a,p_1,\epsilon_1) \leftrightarrow (b,p_2,\epsilon_2)
\end{array}
\eeq
with on--shell conditions
\beq 
\begin{array}{ccl}
p_i \cdot a_i =0 \quad , \quad p_i \cdot p_j =\begin{cases}
-\frac{1}{\alpha'} \,, & i=j \\
+\frac{1}{2\alpha'} \,, & i \neq j\,.
\end{cases}
\end{array}
\eeq
\end{itemize}
In the above, after writing each amplitude as the correlator of the respective vertex operators, we also give the mass level and transversality for each leg, as well as the value of the product $p_i \cdot p_j$
for each case, which is derived by means of the use of the former within (\ref{mom_cons}); we refer collectively to the aforementioned as the on--shell conditions. 

By applying Wick's theorem, we bring the four amplitudes to the schematic form
\beq \label{after_Wick}
\begin{array}{ccl}
\mathcal{A}_{AAA}&=& \frac{g_o^3}{\sqrt{\alpha'}} \, C_{D_2} \, \epsilon_{1\mu} \, \epsilon_{2\nu} \, \epsilon_{3\lambda} \, \Tr([T^a, T^b ]T^c)  \, \mathcal{E}_{AAA}\, \mathcal{B}^{\mu \nu \lambda}
\crbig
\mathcal{A}_{AA\bm{A}} &=&  \frac{g_o^3}{\sqrt{\alpha'}}  \, C_{D_2} \, \epsilon_{1\mu} \, \epsilon_{2\nu} \, a_{3\lambda} \, \Tr([T^a, T^b ]T^c)   \, \mathcal{E}_{AA\bm{A}} \, \mathcal{B}^{\mu \nu \lambda}\, \langle \mathcal{J}_3 \rangle
\crbig
\mathcal{A}_{\bm{A}\bm{A}A} &=&  \frac{g_o^3}{\sqrt{\alpha'}}  \, C_{D_2} \,  a_{1\mu} \, a_{2\nu} \, \epsilon_{3\lambda} \, \Tr([T^a, T^b ]T^c)  \, \mathcal{E}_{\bm{A}\bm{A}A} \, \mathcal{B}^{\mu \nu \lambda} \, \langle  \mathcal{J}_1 \, \mathcal{J}_2 \rangle
\crbig
\mathcal{A}_{\bm{A} \bm{A}\bm{A}}  &=&  \frac{g_o^3}{\sqrt{\alpha'}}  \, C_{D_2} \,  a_{1\mu} \, a_{2\nu} \, a_{3\lambda} \, \Tr([T^a, T^b ]T^c)  \, \mathcal{E}_{\bm{A} \bm{A}\bm{A}} \, \mathcal{B}^{\mu \nu \lambda} \, \langle  \mathcal{J}_1 \, \mathcal{J}_2 \,  \mathcal{J}_3 \rangle  \,,
\end{array}
\eeq
where $a_{i\, \mu}$ is to be understood as $a_{i\, \mu}$, $a_{i\, \mu}^A$ and $a_{i\, \mu}^{MN}$ and $\mathcal{J}_i$ as $\mathcal{J}_i$, $\mathcal{J}_i^A$ and $\mathcal{J}_i^{MN}$ for $\mathcal{N}=1,2,4$ respectively.
Notice that in $\mathcal{A}_{\bm{A} \bm{A}\bm{A}} $ we have omitted a term that schematically takes the form
\beq \label{extra_int_van}
\langle \psi^\mu_1 \psi^\nu_2 \psi^\lambda_3 \rangle \, \langle \mathcal{J}_1 \mathcal{J}_2 \, (\textrm{internal fields})_3 \rangle\,,
\eeq
which originates in the last line of  (\ref{massive_vector_pc_1}), (\ref{massive_vector_pc_2}) and or (\ref{massive_vector_pc_4}) for $\mathcal{N}=1,2,4$ respectively, since $3$--point functions of worldsheet fermions vanish. In the $\mathcal{N}=1$ example, (\ref{extra_int_van}) takes the form
\beq 
\langle \psi^\mu_1 \psi^\nu_2 \psi^\lambda_3 \rangle \, \langle \mathcal{J}_1 \mathcal{J}_2 \,  \big(T_{\textrm{F,int},3}^+ - T_{\textrm{F,int},3}^- \big) \rangle\,.
\eeq
As expected, all amplitudes involve the Koba--Nielsen factor (\ref{KB_open})
\beq \label{KB_open3}
\begin{array}{ccl}
\mathcal{E} \equiv \langle :e^{ip_1X_1}: \, :e^{ip_2X_2}: \, :e^{ip_3X_3}: \rangle \, = z_{12}^{2\alpha' p_1 \cdot p_2} \, z_{13}^{2\alpha' p_3 \cdot p_3} \,z_{23}^{2\alpha' p_2 \cdot p_3}\,,
\end{array}
\eeq
whose  $z$--dependence varies according to the type of the particular amplitude, namely on the value of $p_i \cdot p_j$. We find
\beq \label{E_value_open}
 \mathcal{E}_{AAA}=1\quad , \quad \mathcal{E}_{AA\bm{A}}=\frac{z_{13}z_{23}}{z_{12}} \quad , \quad \mathcal{E}_{\bm{A}\bm{A}A}=z_{12}^2 \quad , \quad \mathcal{E}_{\bm{A}\bm{A}\bm{A}}=z_{12}z_{13}z_{23}\,.
\eeq
Moreover, we observe that within all amplitudes there appears the following product of correlators
\beq \label{univ_corr}
\begin{array}{ccl}
\mathcal{B}^{\mu \nu \lambda} \equiv  \langle c_1 c_2 c_3 \rangle  \langle e^{-\phi_1}e^{-\phi_2} \rangle  \bigg\{ \langle\psi_1^\mu \psi_2^\nu \rangle \Big[ \langle ip_1X_1\, i\partial X_3^\lambda \rangle  + \langle ip_2 X_2\, i\partial X_3^\lambda \rangle  \Big] 
\crbig
-2\alpha' \Big[ \langle \psi_1^\mu \, \big(p_3 \cdot \psi_3\big) \rangle \langle \psi_2^\nu \psi_3^\lambda \rangle - \langle \psi_1^\mu \psi_3^\lambda \rangle \langle \psi_2^\nu \, \big(p_3 \cdot  \psi_3 \big) \rangle \Big] \bigg\} \,.
\crbig
\end{array}
\eeq
Using the disk correlators of the previous section, as well as (\ref{mom_cons}) and the transversality of the external states, we obtain
\beq \label{B_value}
\mathcal{B}^{\mu \nu \lambda} =2\alpha'\, \big( \eta^{\mu \nu} p_1^\lambda +  \eta^{\nu \lambda} p_2^\mu  +  \eta^{\lambda \mu} p_3^\nu \big) \,,
\eeq
which is, therefore, strictly speaking only valid within the amplitudes. Interestingly, the value of (\ref{univ_corr}) is the same for all four amplitudes and does not depend on $z$; the $z$--dependence of the ghost contribution cancels that of the contribution of the ``matter'' fields. We then use (\ref{E_value_open}) and (\ref{B_value}) and of course find
\beq \label{sa_massless_s}
\mathcal{A}_{AAA}^{\mathcal{N}=1,2,4}=\frac{g_o}{\sqrt{\alpha'}}  f^{abc } \,\big( \epsilon_1 \cdot \epsilon_2\, p_1 \cdot \epsilon_3   + \epsilon_2 \cdot \epsilon_3\, p_2 \cdot \epsilon_1  + \epsilon_3 \cdot \epsilon_1\, p_3 \cdot \epsilon_2 \big)\,,
\eeq
where dot products stand for contractions in spacetime indices.

Next we proceed on a case--by--case basis in regard to the amount of brane supersymmetry. For $\mathcal{N}=1,2,4$, we use the Kac--Moody algebras (\ref{km_1}), (\ref{km_2}) and (\ref{km_4}) and obtain in all cases
\beq \label{theorem}
\mathcal{A}_{AA\bm{A}}^{\mathcal{N}=1,2,4} = 0\,.
\eeq
It is worth noticing that this vanishing of the amplitude involving two massless and one massive vector is actually a more general property of field theories: massive particles \textit{cannot} decay into however many particles of the same helicity. This was first observed for massive $U(1)$ vectors by Landau and Yang and was shown to be the case for states of any spin \cite{Arkani-Hamed:2017jhn}. We see here that for the vectors that we consider in compactified string theory, it is due to the appearance of the $1$--point function of the Kac--Moody current. We further find for all cases $\mathcal{N}=1,2,4$
\beq \label{same_ampl}
\mathcal{A}_{\bm{A}\bm{A}A}^{\mathcal{N}=1,2,4}  = \frac{g_o}{\sqrt{\alpha'}} \, f^{abc }\,\big( a_1 \cdot a_2\, p_1 \cdot \epsilon_3   + a_2 \cdot \epsilon_3\, p_2 \cdot a_1  + \epsilon_3 \cdot a_1\, p_3 \cdot a_2 \big) 
\eeq
where in $\mathcal{A}_{\bm{A}\bm{A}A}^{\mathcal{N}=2}$ the Kac--Moody $SU(2)$ indices of $a_1^A$ and $a_2^B$ are implicitly contracted with each other, while in $\mathcal{A}_{\bm{A}\bm{A}A}^{\mathcal{N}=4}$ the Kac--Moody $SO(6)$ are contracted, namely $a_1 \cdot a_2$ stands for $a_1^{MN} \cdot a_2^{MN}$; recall that $a_1^{MN}$ is antisymmetric w.r.t. the interchange of $M$ and $N$. 

Finally, we obtain
\beq \label{ampl_vec_1}
\mathcal{A}_{\bm{A}\bm{A}\bm{A}}^{\mathcal{N}=1}  = 0\,,
\eeq
where we highlight that the reason behind the vanishing is that $3$--point functions of $U(1)$ worldsheet currents vanish. For $\mathcal{N}=2$ however, we obtain
\beq \label{ampl_vec_2}
\mathcal{A}_{\bm{A}\bm{A}\bm{A}}^{\mathcal{N}=2}  =\frac{g_o}{\sqrt{\alpha'}}  \, f^{abc } \varepsilon^{ABC} \,\big( a_1^A \cdot a_2^B \, p_1 \cdot a_3^C   + a_2^B \cdot a_3^C \, p_2 \cdot a_1^A  + a_3^C \cdot a_1^A \, p_3 \cdot a_2^B \big)\,.
\eeq
 Notice that $\mathcal{A}_{\bm{A}\bm{A}\bm{A}}$ is now generically non--zero, since $f^{abc } \varepsilon^{ABC}$ is \textit{symmetric} under exchange of legs $1$ and $2$ for example. For $\mathcal{N}=4$, we also obtain a non--vanishing result
\beq \label{ampl_vec_4}
\mathcal{A}_{\bm{A}\bm{A}\bm{A}}^{\mathcal{N}=4} = \frac{g_o}{\sqrt{\alpha'}} \, f^{abc } \,\big( a_1^{MN} \cdot a_2^{ML} \, p_1 \cdot a_3^{NL}   + a_2^{ML} \cdot a_3^{NL} \, p_2 \cdot a_1^{MN}  + a_3^{NL} \cdot a_1^{MN} \, p_3 \cdot a_2^{ML} \big)\,.
\eeq
In all these cases, the varying $z$--dependence of the Koba--Nielsen function is compensated for by means of the varying contribution of the Kac--Moody current correlators, as more legs are taken to be massive. Notice further that, as expected, all amplitudes are symmetric under cyclic interchanges of the three external legs.

At this point, it is instructive to compare with the bosonic case. Using the massless (\ref{gaugevo_open_b_comp}) and massive (\ref{massive_vector_b}) vertex operators, as well as the Kac--Moody algebra (\ref{km_bosonic}), we find
\begin{subequations}
\begin{align}
\mathcal{A}_{AAA}^{\textrm{bos}}&= \frac{g_o}{\sqrt{\alpha'}}   f^{abc } \,\big( \epsilon_1 \cdot \epsilon_2\, p_1 \cdot \epsilon_3   + \epsilon_2 \cdot \epsilon_3\, p_2 \cdot \epsilon_1  + \epsilon_3 \cdot \epsilon_1\, p_3 \cdot \epsilon_2\,,  \nonumber
\crbig
& \qquad + 2\alpha' \, \epsilon_1 \cdot p_2 \, \epsilon_2 \cdot p_3 \, \epsilon_3 \cdot p_1  \big) \label{vec_ampl_1_b}
\crbig
\mathcal{A}_{AA\bm{A}}^{\textrm{bos}} &= 0  \label{vec_ampl_2_b}
\crbig
\mathcal{A}_{\bm{A}\bm{A}A}^{\textrm{bos}}  &= \frac{g_o}{\sqrt{\alpha'}}  \, f^{abc }\,\big( a_1 \cdot a_2\, p_1 \cdot \epsilon_3   + a_2 \cdot \epsilon_3\, p_2 \cdot a_1  + \epsilon_3 \cdot a_1\, p_3 \cdot a_2 \nonumber
\crbig
& \qquad + 2\alpha' \, a_1 \cdot p_2 \, a_2 \cdot p_3 \, \epsilon_3 \cdot p_1   \big)  \label{vec_ampl_3_b}
\crbig
\mathcal{A}_{\bm{A}\bm{A}\bm{A}}^{\textrm{bos}}  &= 0\,, \label{vec_ampl_4_b}
\end{align}
\end{subequations}
where  in $\mathcal{A}_{\bm{A}\bm{A}A}^{\textrm{bos}}$ the Kac--Moody indices of $a_1^M$ and $a_2^N$ are implicitly contracted with each other. Notice that $\mathcal{A}_{AA\bm{A}}^{\textrm{bos}}$ and $\mathcal{A}_{\bm{A}\bm{A}\bm{A}}^{\textrm{bos}} $ vanish because $1$-- and $3$--point functions of the currents in question vanish. The situation thus resembles most that of $\mathcal{N}=1$, up to the $3$--momentum term that is forbidden by supersymmetry and which in the bosonic case appears at one order in $\alpha'$ higher than the rest, as is also the case before compactifications.

\subsection{Spin--$2$ scattering}

In this subsection, each external leg in the $(-1,-1)$ picture is associated with the vertex operator (\ref{graviton_vo}) if massless and with (\ref{dc_massive}) if massive and each in the $(0,0)$ picture with (\ref{dc_zero}). We can thus write schematically
\begin{itemize}
\item three gravitons: to the best of our knowledge, this is the only case that has been studied before cf. e.g. \cite{Gross:1986mw, Lust:1987xm}
\beq \label{M3_massless}
\begin{array}{ccl}
\mathcal{M}_{GGG} = \langle  V_{G}^{(-1,-1)}(z_1,\varepsilon_1,k_1) \,  V_{G}^{(-1,-1)}(z_2,\varepsilon_2,k_2)  \,V_{G}^{(0,0)}(z_3,\varepsilon_3,k_3) \rangle
\crbig
 \times \langle c_1 \widetilde{c}_1  \, c_2 \widetilde{c}_2  \, c_3 \widetilde{c}_3  \rangle
\crbig
k_i^2 =0 \quad , \quad k_i^\mu \varepsilon_{i\, \mu \nu}=0 \quad, \quad k_i \cdot k_j =0 \,,
\end{array}
\eeq
\item two gravitons and one massive spin--$2$ state:
\beq \label{M3_massless_massive1}
\begin{array}{ccl}
\mathcal{M}_{GGM}  =\langle  V_{G}^{(-1,-1)}(z_1,\varepsilon_1,k_1) \,  V_{G}^{(-1,-1)}(z_2,\varepsilon_2,k_2)  \,V_{M}^{(0,0)}(z_3,\alpha,k_3) \rangle
\crbig
 \times \langle c_1 \widetilde{c}_1  \, c_2 \widetilde{c}_2  \, c_3 \widetilde{c}_3  \rangle
\crbig
k_1^\mu  \varepsilon_{1\,\mu\nu}= k_2^\mu  \varepsilon_{2\, \mu \nu} = k_3^\mu  \alpha_{\mu \nu}=0
\crbig
k_1^2=k_2^2=0 \quad , \quad k_3^2 =-\frac{4}{\alpha'} \quad , \quad k_1 \cdot k_2 = -\frac{2}{\alpha'} \quad , \quad k_1 \cdot k_3 = k_2 \cdot k_3=\frac{2}{\alpha'}\,,
\end{array}
\eeq
\item two massive spin--$2$ states and one graviton:
\beq \label{M3_massless_massive2}
\begin{array}{ccl}
\mathcal{M}_{MMG}  =\langle  V_{M}^{(-1,-1)}(z_1,\alpha_1,k_1) \, V_{M}^{(-1,-1)}(z_2,\alpha_2,k_2) \,V_{G}^{(0,0)}(z_3,\varepsilon,k_3) \rangle
\crbig
 \times \langle c_1 \widetilde{c}_1  \, c_2 \widetilde{c}_2  \, c_3 \widetilde{c}_3  \rangle
\crbig
k_1^\mu  \alpha_{1\,\mu \nu}= k_2^\mu \alpha_{2\,\mu \nu} = k_3^\mu \varepsilon_{\mu \nu}=0
\crbig
k_1^2=k_2^2 =-\frac{4}{\alpha'}  \quad , \quad k_3^2=0 \quad , \quad k_1 \cdot k_2 = \frac{4}{\alpha'} \quad , \quad k_1 \cdot k_3 = k_2 \cdot k_3=0\,,
\end{array}
\eeq
\item three massive spin--$2$ states:
\beq \label{M3_massive}
\begin{array}{ccl}
\mathcal{M}_{MMM}  =\langle V_{M}^{(-1,-1)}(z_1,\alpha_1,k_1) \,  V_{M}^{(-1,-1)}(z_2,\alpha_2,k_2) \, V_{M}^{(0,0)}(z_3,\alpha_3,k_3) \rangle
\crbig
 \times \langle c_1 \widetilde{c}_1  \, c_2 \widetilde{c}_2  \, c_3 \widetilde{c}_3  \rangle
\crbig
k_i^\mu  \alpha_{i\, \mu \nu} =0 \quad , \quad k_i \cdot k_j =\begin{cases}
-\frac{4}{\alpha'} \,, & i=j \\
+\frac{2}{\alpha'} \,, & i \neq j\,.
\end{cases}
\end{array}
\eeq

\end{itemize}

We next exploit the property of the spin--2 states of being formulated as products of vectors, as well as our results of the previous subsection and arrive at
\beq
\begin{array}{ccl}
\mathcal{M}_{GGG} &=&\frac{g_{\textrm{c}}^3}{\alpha'} \, C_{S_2} \,   \varepsilon_{1\mu \rho} \, \varepsilon_{2\nu \sigma} \, \varepsilon_{3\lambda \kappa} \,E_{GGG}\, \mathcal{B}^{\mu \nu \lambda} \,\widetilde{\mathcal{B}}^{\rho \sigma \kappa}
\crbig
\mathcal{M}_{GGM}&=& \frac{g_{\textrm{c}}^3}{\alpha'} \, C_{S_2} \,   \varepsilon_{1\mu \rho} \, \varepsilon_{2\nu \sigma} \, \alpha_{\lambda \kappa} \,E_{GGM} \, \mathcal{B}^{\mu \nu \lambda}\, \widetilde{\mathcal{B}}^{\rho \sigma \kappa}\, \langle \mathcal{J}_3 \rangle \, \langle \widetilde{\mathcal{J}}_3 \rangle
\crbig
\mathcal{M}_{MMG} &=& \frac{g_{\textrm{c}}^3}{\alpha'} \, C_{S_2} \,   \alpha_{1\mu \rho} \, \alpha_{2\nu \sigma} \, \varepsilon_{\lambda \kappa} \, E_{MMG} \, \mathcal{B}^{\mu \nu \lambda} \, \widetilde{\mathcal{B}}^{\rho \sigma \kappa}\, \langle \mathcal{J}_1 \, \mathcal{J}_2 \rangle\, \langle \widetilde{\mathcal{J}}_1 \, \widetilde{\mathcal{J}}_2 \rangle
\crbig
\mathcal{M}_{MMM} &=& \frac{g_{\textrm{c}}^3}{\alpha'} \, C_{S_2} \,    \alpha_{1\mu \rho} \, \alpha_{2\nu \sigma} \, \alpha_{3\lambda \kappa} \, E_{MMM} \, \mathcal{B}^{\mu \nu \lambda} \, \widetilde{\mathcal{B}}^{\rho \sigma \kappa}\, \langle \mathcal{J}_1 \, \mathcal{J}_2 \, \mathcal{J}_3 \rangle \, \langle \widetilde{\mathcal{J}}_1 \, \widetilde{\mathcal{J}}_2  \widetilde{\mathcal{J}}_3 \rangle
\end{array}
\eeq
where $\alpha_{i\, \mu \nu}$ is to be understood as $\alpha_{i\, \mu \nu}$, $\alpha_{i\, \mu \nu}^{AA'}$ and $\alpha_{i\, \mu \nu}^{MNM'N'}$ and $\mathcal{J}_i$ as $\mathcal{J}_i$, $\mathcal{J}_i^A$ and $\mathcal{J}_i^{MN}$ (similarly for its antiholomorphic counterparts) for $\mathcal{N}=2,4,8$ respectively. As expected, all amplitudes involve the Koba--Nielsen factor 
\beq \label{KB_closed3}
\begin{array}{ccl}
E \equiv \langle :e^{ik_1X(z_1, \ov{z}_1}: \, :e^{ik_2X(z_2, \ov{z}_2)}: \, :e^{ik_3X(z_3, \ov{z}_3)}: \rangle \, = |z_{12}|^{\alpha' k_1 \cdot k_2} \, |z_{13}|^{\alpha' k_3 \cdot k_3} \, |z_{23}|^{\alpha' k_2 \cdot k_3}\,,
\end{array}
\eeq
it is easy to see that
\beq
E= \mathcal{E} \widetilde{\mathcal{E}}
\eeq
and that
\beq \label{E_value_closed}
\begin{array}{ccl}
E_{GGG}=1\quad , \quad E_{GGM}=\frac{z_{13}z_{23}}{z_{12}} \frac{\ov{z}_{13} \ov{z}_{23}}{\ov{z}_{12}}  
\crbig
E_{MMG}=z_{12}^2 \ov{z}_{12}^2 \quad , \quad E_{MMM}=z_{12}z_{13}z_{23} \, \ov{z}_{12}\ov{z}_{13}\ov{z}_{23}\,.
\end{array}
\eeq
Consequently, it is useful to define the colour--ordered 3--point vector amplitudes
\beq
\begin{array}{ccl}
\ov{\mathcal{A}}_{AAA} &\equiv& \frac{g_{\textrm{o}}^3}{\sqrt{\alpha'}} \, \epsilon_{1\mu} \, \epsilon_{2\nu} \, \epsilon_{3\lambda} \, \mathcal{E}_{AAA}\, \mathcal{B}^{\mu \nu \lambda}
\crbig
\ov{\mathcal{A}}_{AA\bm{A}}&\equiv& \frac{g_{\textrm{o}}^3}{\sqrt{\alpha'}} \, \epsilon_{1\mu} \, \epsilon_{2\nu} \, a_{3\lambda} \, \mathcal{E}_{AA\bm{A}} \, \mathcal{B}^{\mu \nu \lambda}\, \langle \mathcal{J}_3 \rangle
\crbig
\ov{\mathcal{A}}_{\bm{A}\bm{A}A} &\equiv& \frac{g_{\textrm{o}}^3}{\sqrt{\alpha'}}  \, a_{1\mu} \, a_{2\nu} \, \epsilon_{3\lambda} \,  \mathcal{E}_{\bm{A}\bm{A}A} \, \mathcal{B}^{\mu \nu \lambda} \, \langle \mathcal{J}_1 \, \mathcal{J}_2 \rangle
\crbig
\ov{\mathcal{A}}_{\bm{A}\bm{A}\bm{A}} &\equiv& \frac{g_{\textrm{o}}^3}{\sqrt{\alpha'}}  \,  a_{1\mu} \, a_{2\nu} \, a_{3\lambda} \,  \mathcal{E}_{\bm{A} \bm{A}\bm{A}} \, \mathcal{B}^{\mu \nu \lambda} \, \langle \mathcal{J}_1 \, \mathcal{J}_2 \, \mathcal{J}_3 \rangle\,,
\end{array}
\eeq
as well as their antiholomorphic counterparts w.r.t. $\widetilde \epsilon_{i \, \mu}$ and $\widetilde a_{i\, \mu}$. We highlight that by ``colour--ordered'' we refer to the \textit{Chan--Paton} gauge algebras, with the Kac--Moody left intact. We then obtain a manifestly \textit{double copy} structure
\beq
\begin{array}{ccl}
\mathcal{M}_{GGG}=\ov{\mathcal{A}}_{AAA} \cdot \widetilde{\ov{\mathcal{A}}}_{AAA} \quad, \quad \mathcal{M}_{GGM} = \ov{\mathcal{A}}_{AA\bm{A}}  \cdot \widetilde{\ov{\mathcal{A}}}_{AA\bm{A}}
\crbig
\mathcal{M}_{MMM} =   \ov{\mathcal{A}}_{\bm{A}\bm{A}\bm{A}}   \cdot \widetilde{\ov{\mathcal{A}}}_{\bm{A}\bm{A}\bm{A}} \quad , \quad \mathcal{M}_{MMG} = \ov{\mathcal{A}}_{\bm{A}\bm{A}A}   \cdot \widetilde{\ov{\mathcal{A}}}_{\bm{A}\bm{A}A} 
\end{array}
\eeq
upon assuming
\beq
\varepsilon_{i \, \mu \nu} = \epsilon_{i \, \mu} \widetilde{\epsilon}_{i \, \nu} \quad , \quad \alpha_{i \, \mu \nu} = a_{i \, \mu} \widetilde{a}_{i \, \nu} 
\eeq
and of course that $\varepsilon_{i \, \mu \nu}$, $a_{i \, \mu \nu}$ be symmetric and traceless, or equivalently (\ref{pol_G_prod}) and (\ref{massive_pol}).  

Using the above, we have
\beq \label{univ_g}
\begin{array}{ccl}
\mathcal{M}_{GGG}^{\mathcal{N}=2,4,8}&=& g_{\textrm{c}}\Big[   (k_1 \cdot \varepsilon_3 \cdot  k_1) \Tr\big(\varepsilon_1 \cdot \varepsilon_2\big)+ 2 \, k_2 \cdot \varepsilon_1 \cdot \varepsilon_2 \cdot \varepsilon_3 \cdot k_1 \Big]
\crbig
&& \qquad + \, \textrm{cyclic permutations} 
\end{array}
\eeq
as expected. Next, we obtain
\beq \label{van_m_s}
\begin{array}{ccl}
\mathcal{M}_{GGM}^{\mathcal{N}=2,4,8} &=&0\,,
\end{array}
\eeq
we namely show that our massive spin--2 states cannot decay into massless gravitons (at least to cubic order), thus observing accordance with the \textit{field theory} conclusions of \cite{Arkani-Hamed:2017jhn} also for spin--$2$. In addition, we find that
\beq \label{one_massless_same}
\begin{array}{ccl}
\mathcal{M}_{MMG}^{\mathcal{N}=2,4,8}&=& g_{\textrm{c}}\Big[ (k_1 \cdot \varepsilon \cdot k_1) \Tr(\alpha_1 \cdot \alpha_2) + 2 \, k_2 \cdot \alpha_1 \cdot \alpha_2 \cdot \varepsilon \cdot k_1 \Big]\,,
\crbig
&& \qquad + \, \textrm{cyclic permutations} 
\end{array}
\eeq
where in $\mathcal{M}_{MMG}^{\mathcal{N}=4}$ the Kac--Moody $SU(2)$ indices of $\alpha_1^{AA'}$ and $\alpha_2^{BB'}$ are implicitly contracted with each other, while in $\mathcal{M}_{MMG}^{\mathcal{N}=8}$ those of the Kac--Moody $SO(6)$ are contracted, for example
\beq
(k_1 \cdot \alpha_2 \cdot k_1) \Tr (\alpha_1 \cdot \varepsilon)  = \begin{cases}
(k_1 \cdot \alpha_2^{AA'} \cdot k_1) \Tr (\alpha_1^{AA'} \cdot \varepsilon) \\
 (k_1 \cdot \alpha_2^{MNM'N'} \cdot k_1) \Tr (\alpha_1^{MNM'N'} \cdot \varepsilon)\,.
\end{cases}
\eeq
Finally, we obtain
\beq \label{one_van_ampl}
\mathcal{M}_{MMM}^{\mathcal{N}=2} = 0\,,
\eeq
again due to the vanishing of $3$--point functions of worldsheet currents,
\beq \label{three_m_4}
\begin{array}{ccl}
\mathcal{M}_{MMM}^{\mathcal{N}=4}&=&  g_{\textrm{c}}  \varepsilon^{ABC} \varepsilon^{A'B'C'} \,\Big[  (k_1 \cdot \alpha_3^{CC'} \cdot k_1) \Tr(\alpha_1^{AA'} \cdot \alpha_2^{BB'})
\crbig
&& \qquad + 2 \, k_2 \cdot \alpha_1^{AA'} \cdot \alpha_2^{BB'} \cdot \alpha_3^{CC'} \cdot k_1 \Big]
\crbig
&& \qquad + \, \textrm{cyclic permutations} 
\end{array}
\eeq
and
\beq \label{three_m_8}
\begin{array}{ccl}
\mathcal{M}_{MMM}^{\mathcal{N}=8}&=&  g_{\textrm{c}} \, \Big[  (k_1 \cdot \alpha_3^{NLN'L'} \cdot k_1) \Tr(\alpha_1^{MNM'N'} \cdot \alpha_2^{MLM'L'})
\crbig
&&\qquad + 2 \, k_2 \cdot \alpha_1^{MNM'N'} \cdot \alpha_2^{MLM'L'} \cdot \alpha_3^{NLN'L'} \cdot k_1 \Big]
\crbig
&& \qquad + \, \textrm{cyclic permutations}  \,.
\end{array}
\eeq
As expected, $\mathcal{M}_{GGG}$ and $\mathcal{M}_{MMM}$ are symmetric under cyclic interchanges of the three external legs and $\mathcal{M}_{MMG}$ is further symmetric upon exchanging the two massive legs.

At this point, it is important to recall that our massive spin--2 states of $\mathcal{N}=2,4,8$ transform in non--trivial representations of the respective $R$--symmetry. Nevertheless, later in this work we will be comparing our amplitudes results with theories of massive spin--2 states that carry no nonabelian charges. Consequently, we now select by hand the singlet representation in each case and stress that this choice does \textit{not} preserve the full bulk supersymmetry. This implies that in the following we must think of our massive spin--2 states as \textit{non--BPS} states and, to keep track of this property, we will denote these cases with $\mathcal{N}=\widetilde{2}, \widetilde{4}, \widetilde{8}$. Then (\ref{three_m_4}) and (\ref{three_m_8}) for example both boil down to
\beq \label{massive_3_fin}
\begin{array}{ccl}
\mathcal{M}_{MMM}^{\mathcal{N}=\widetilde 4, \widetilde 8}&=&  g_{\textrm{c}}  \Big[  (k_1 \cdot \alpha_3 \cdot k_1) \Tr(\alpha_1 \cdot \alpha_2) + 2 \, k_2 \cdot \alpha_1 \cdot \alpha_2 \cdot \alpha_3 \cdot k_1 \Big]
\crbig
&& \qquad + \, \textrm{cyclic permutations}  \,.
\end{array}
\eeq
Interestingly, comparison of (\ref{univ_g}) with (\ref{massive_3_fin}) reveals that the $3$--point amplitude of three identical states is the \textit{same} regardless of whether these are gravitons or the massive spin--$2$ states from $\mathcal{N}=\widetilde{4}, \widetilde{8}.$

Finally, let us compare with the case of the bosonic string. Using our results of the previous subsection, it is straightforward to find
\begin{subequations}
\begin{align}
\mathcal{M}_{GGG}^{\textrm{bos}}&= g_{\textrm{c}}\Big[   (k_1 \cdot \varepsilon_3 \cdot  k_1) \Tr\big(\varepsilon_1 \cdot \varepsilon_2\big)+ 2 \, k_2 \cdot \varepsilon_1 \cdot \varepsilon_2 \cdot \varepsilon_3 \cdot k_1  \nonumber
\crbig
& \qquad  +\alpha'  \,  (k_1 \cdot \varepsilon_3 \cdot  k_1) \, (k_2 \cdot \varepsilon_1 \cdot \varepsilon_2 \cdot  k_3) \nonumber
\crbig
& \qquad + \frac{\alpha'^2}{12} \,  (k_2 \cdot \varepsilon_1 \cdot  k_2)   (k_3 \cdot \varepsilon_2 \cdot  k_3)   (k_1 \cdot \varepsilon_3 \cdot  k_1)  \Big]  \nonumber  
\crbig
& \qquad  + \textrm{cyclic permutations} \label{ampl_2_bosonic_1}
\crbig
 \mathcal{M}_{GGM}^{\textrm{bos}} &=0\,, \label{ampl_2_bosonic_2}
 \crbig
 \mathcal{M}_{MMG}^{\textrm{bos}}&= g_{\textrm{c}}\Big[ (k_1 \cdot \varepsilon \cdot k_1) \Tr(\alpha_1 \cdot \alpha_2)  + 2 \, k_2 \cdot \alpha_1 \cdot \alpha_2 \cdot \varepsilon \cdot k_1 \nonumber
 \crbig
& \qquad  +\alpha'  \,  (k_1 \cdot \varepsilon \cdot  k_1) \, (k_2 \cdot \alpha_1 \cdot \alpha_2 \cdot  k_3) \nonumber
\crbig
& \qquad + \frac{\alpha'^2}{12} \,  (k_2 \cdot \alpha_1 \cdot  k_2)   (k_3 \cdot \alpha_2 \cdot  k_3)   (k_1 \cdot \varepsilon \cdot  k_1)  \Big]  \nonumber  
\crbig
& \qquad  + \textrm{cyclic permutations}  \label{ampl_2_bosonic_3}
\crbig 
\mathcal{M}_{MMM}^{\textrm{bos}}&= 0\,, \label{ampl_2_bosonic_4}
 \end{align}
\end{subequations}
namely agreement with our $\mathcal{N}=2$ results to lowest order in $\alpha'$, as again $4$-- and $6$--momenta terms are not allowed by supersymmetry.

Before concluding this section, let us investigate how different our non--vanishing cubic string amplitudes from those of field theory are. In the string side, we have found that the amplitude of one graviton and two massive spin--2 (\ref{one_massless_same}) (for all cases $\mathcal{N}=2,4,8$) and  the 3--massive amplitude (\ref{massive_3_fin}) (for $\mathcal{N}=4,8$) involve solely two kinds, namely disparate Lorentz contractions, of $2$--momenta terms; these are all possible $2$--momenta terms for on--shell, transverse and traceless spin--2 states. Upon considering the basis of linearly independent cubic amplitudes with gravitons and massive spin--2 fields as external states constructed in \cite{Hinterbichler:2017qyt, Bonifacio:2017nnt}, we see that both kinds of $2$--momenta terms are allowed also in the field theory side for both amplitudes, albeit with their relative coefficients interestingly \textit{a priori not} fixed to a number, unlike our string results (\ref{one_massless_same}), (\ref{massive_3_fin}). The physical significance of these relative coefficients we will explore in the next section. We also note that further terms of higher order in momenta are not forbidden \cite{Hinterbichler:2017qyt, Bonifacio:2017nnt} in the field theory side, similarly to our bosonic result (\ref{ampl_2_bosonic_3}). 

\section{``Effective'' Lagrangian theory} \label{vertices}

\subsection{From 3--point amplitudes to cubic vertices}

Extracting low--energy effective actions from string theory is well understood for massless external states: one replaces polarisation tensors with fields and momenta with derivatives in the amplitudes' expressions and takes the limit $\alpha' \rightarrow 0$ to suppress contributions from the tower of massive states in the internal lines of the respective Fenyman diagrams. From our cubic amplitudes with massless external legs, evidently no such limit needs to be taken, so simply by replacing (in four spacetime dimensions)
\beq \label{repl1}
\epsilon_\mu \rightarrow A_\mu^a  \quad , \quad p^\mu \rightarrow i \partial^\mu \quad 
\eeq
in (\ref{sa_massless_s}) we obtain the known result
\beq \label{known_1}
\begin{array}{ccl}
 \mathcal{L}_{A^3}^{\mathcal{N}=1,2,4}&=&\frac{g_{\textrm{o}}} {\sqrt{\alpha'}}\, f^{abc} \, \big( \partial^\mu A_\nu^a \big) \, A^{\nu \, b} A_\mu^c
 \end{array}
\eeq
and by replacing 
\beq \label{repl2}
 \varepsilon_{\mu \nu} \rightarrow G_{\mu \nu} \quad , \quad k^\mu \rightarrow i \partial^\mu
\eeq
in (\ref{univ_g}) we obtain the known result
\beq \label{known_2}
\begin{array}{ccl}
 \mathcal{L}_{\textrm{G}^3}^{\mathcal{N}=2,4,8}&=&g_{\textrm{c}} \, G^{\mu \nu}\big(\partial_\mu  G_{\rho \sigma} \partial_\nu G^{\rho \sigma}  -2\partial_\nu  G_{\rho \sigma} \partial^\sigma  G_\mu^\rho\big)  \,.    
\end{array}
\eeq
The bosonic cases (\ref{vec_ampl_1_b}) and in (\ref{ampl_2_bosonic_1}) yield the very same results of course up to $\mathcal{F}^3$ and $\mathcal{R}^3$ terms, with $\mathcal{F}_{\mu \nu}$ the Yang--Mills field--strength and $\mathcal{R}_{\mu \nu \kappa \lambda}$ the Riemann tensor, that are not allowed by supersymmetry. 

Turning to amplitudes with massive external legs: a complete Lagrangian description derived as an effective low--energy theory from scattering amplitudes carries no meaning, as the infinitely many massive string states all become relevant at the same energy scale, namely $\alpha'$, as highlighted in section \ref{intro}. Nevertheless, we observe that all our cubic results of the previous section are \textit{finite} and not expansions in $\alpha'$. We would thus like to examine the kind of cubic vertices they correspond to. In particular, we will be employing (\ref{repl1}), (\ref{repl2}) and
\beq
a_\mu \rightarrow \bm{A}_\mu^a \quad , \quad \alpha_{\mu \nu}  \rightarrow M_{\mu \nu}  \,.
\eeq
From (\ref{theorem}) we find
\beq \label{new_1a}
\begin{array}{ccl}
 \mathcal{L}_{A^2 \bm{A}}^{\mathcal{N}=1,2,4}= 0 
\end{array}
\eeq
while from (\ref{van_m_s})
\beq \label{new_1b}
\begin{array}{ccl}
  \mathcal{L}_{\textrm{G}^2\textrm{M}}^{\mathcal{N}=2,4,8}= 0\,,
\end{array}
\eeq
there are namely \textit{no} cubic vertices involving one massive and two massless states of the same spin. The bosonic cases (\ref{vec_ampl_2_b}) and (\ref{ampl_2_bosonic_2}) yield the very same results. Next, from (\ref{same_ampl}) we obtain 
\beq \label{new_2a}
\begin{array}{ccl}
 \mathcal{L}_{\bm{A}^2A}^{\mathcal{N}=1,2,4}&=&\frac{g_o}{\sqrt{\alpha'}} \, f^{abc} \,  \Big[ \big( \partial^\mu \bm{A}_\nu^a \big) \, \bm{A}^{\nu \, b} A_\mu^c +2 \bm{A}_\mu^a \,\big( \partial^\mu \bm{A}_\nu^b \big) \, A^{\nu \, c}  \Big]
\end{array}
\eeq
and from (\ref{one_massless_same})
\beq \label{new_2b}
\begin{array}{ccl}
  \mathcal{L}_{\textrm{GM}^2}^{\mathcal{N}=2,4,8}&=&g_c \, \Big[ G^{\mu\nu}\big(\partial_\mu M_{\rho \sigma}  \partial_\nu M^{\rho \sigma}  -4 \partial_\nu M_{\rho \sigma}  \partial^\sigma M_{\mu}^{\rho}   \big)
\crbig
&& \qquad \qquad + 2 M^{\mu\nu} \big( \partial_\mu G_{\rho \sigma}  \partial_\nu M^{\rho \sigma}-\partial_\rho G_{\mu \sigma}  \partial_\nu M^{\rho \sigma} \big) \Big]\,.
\end{array}
\eeq
The bosonic cases (\ref{vec_ampl_3_b}) and (\ref{ampl_2_bosonic_3}) yield the very same results. Notice that up to now, the cubic interactions of all the massive spin--2 states we have considered (and of all the massive vectors), for any amount of supersymmetry preserved, are \textit{indistinguishable} from each other.

We now turn to 3--massive amplitudes. From (\ref{ampl_vec_1}) and (\ref{vec_ampl_4_b}) we obtain
\beq \label{case_1a}
 \mathcal{L}_{A^3}^{\mathcal{N}=1} = 0 \quad , \quad  \mathcal{L}_{A^3}^{\textrm{bos}} = 0
\eeq
and from (\ref{one_van_ampl}) and (\ref{ampl_2_bosonic_4})
\beq \label{case_1b}
 \mathcal{L}_{\textrm{M}^3}^{\mathcal{N}=2} = 0 \quad , \quad   \mathcal{L}_{\textrm{M}^3}^{\textrm{bos}} = 0.
\eeq
Next, from (\ref{ampl_vec_2}) and (\ref{ampl_vec_4}) we find
\beq \label{case_2a}
\begin{array}{ccl}
 \mathcal{L}_{A^3}^{\mathcal{N}=2} &=& \frac{g_o}{\sqrt{\alpha'}} \, f^{abc} \varepsilon^{ABC} \,  \big( \partial^\mu \bm{A}_\nu^{aA} \big) \, \bm{A}^{\nu \, bB} \bm{A}_\mu^{cC}
\crbig
 \mathcal{L}_{A^3}^{\mathcal{N}=4} &=& \frac{g_o}{\sqrt{\alpha'}}  \, f^{abc}  \big( \partial^\mu \bm{A}_\nu^{aMN} \big) \, \bm{A}^{\nu \, bML} \bm{A}_\mu^{cNL}

\end{array}
\eeq
and from (\ref{massive_3_fin}) we obtain 
\beq \label{effective_L}
\begin{array}{ccl}
 \mathcal{L}_{\textrm{M}^3}^{\mathcal{N}=\widetilde 4, \widetilde 8} &=& g_c \, M^{\mu \nu}\big(\partial_\mu  M_{\rho \sigma} \partial_\nu M^{\rho \sigma}  -2\partial_\nu  M_{\rho \sigma} \partial^\sigma  M_\mu^\rho\big)    \,.
\end{array}
\eeq
Crucially, the cubic self--interactions (\ref{effective_L}) of the lowest--lying massive spin--2 bulk states we are considering, that originate in multiplets of $\mathcal{N}=4,8$ bulk supersymmetry but do not fully preserve the latter, are thus \textit{identical} to those (\ref{known_2}) of the graviton. Since the structure of the graviton's self--interactions is uniquely determined by the Einstein--Hilbert action and associated with spacetime diffeomorphism invariance, a priori we do not expect this similarity to be maintained at the level of higher--point interactions.

\subsection{Comparison with ghost--free bimetric theory}

Our cubic results describe the interactions of massive spin--2 ``fields'' with the graviton in four spacetime dimensions, as extracted from string theory. We would now like to compare these vertices with an existing field theory that propagates precisely these degrees of freedom around maximally symmetric spacetimes, namely ghost--free bimetric theory. More specifically, the Hassan--Rosen action without matter sources reads \cite{Hassan:2011tf}
\beq \label{HR}
S= m_g^2 \int \md^4x \,\sqrt{g} R(g) +  \alpha^2 m_g^2 \int \md^4x \,\sqrt{f} R(f)  - 2\, \alpha^2 m_g^4  \int \md^4x \,\sqrt{g}\, V(S;\beta_n) \,,
\eeq
where $g_{\mu \nu}$ and $f_{\mu \nu}$ are two distinct rank--2 tensors and $m_g$ and $\alpha m_g$ their respective Planck masses. They interact via the non--derivative potential
\beq \label{pot}
V(S;\beta_n) =  \sum\limits_{n=0}^4 \beta_n e_n(S) \,,
\eeq
where $\beta_n$ are five a priori free dimensionless parameters and $e_n$ elementary symmetric polynomials that can be defined via
\beq 
e_n(S)= S^{\mu_1}_{\hphantom{\mu_1} [\mu_1} \dots S^{\mu_n}_{\hphantom{\mu_n} \mu_n]} \,.
\eeq
Their argument is the matrix $S^\mu_{\hphantom{\mu}\nu}$, that is defined via
\beq \label{square_root}
S^\mu_{\hphantom{\mu}\nu} S^\nu_{\hphantom{\mu}\rho} = g^{\mu \sigma} f_{\sigma \rho} \quad \textrm{namely}  \quad S^\mu_{\hphantom{\mu}\nu} = (\sqrt{g^{-1}f})^\mu_{\hphantom{\mu}\nu}   \,. 
\eeq
We refer the reader to \cite{Lust:2021jps} for a selection of comments on ghost--free bimetric theory that are related to the present investigation, as well as for an indicative list of further references. 

As also highlighted in \cite{Lust:2021jps}, (\ref{HR}) generically propagates $7$ d.o.f that do not generically split into those of the graviton and of the massive spin--2. The simplest way \cite{Hassan:2011tf, Hassan:2012wr} to obtain a well--defined mass spectrum is to assume Minkowski background values for both $g_{\mu \nu}$ and $f_{\mu \nu}$, in which case linear perturbations according to 
\beq
    g_{\mu\nu}=\eta_{\mu\nu}+ \delta g_{\mu\nu} \quad , \quad  f_{\mu\nu}= \eta_{\mu\nu}+ \delta f_{\mu\nu}\,.
\eeq
can be diagonalised into two spin--2 mass eigenstates,
\beq \label{eigen}
 G_{\mu \nu}\equiv m_g \, (\delta g_{\mu \nu} + \alpha^2 \delta f_{\mu \nu}) \quad ,  \quad  M_{\mu \nu} \equiv \alpha  m_g \, (\delta f_{\mu\nu} - \delta g_{\mu \nu})\,,
\eeq
a massless $G_{\mu \nu}$ and a massive $M_{\mu \nu}$ one, with the Fierz--Pauli mass\footnote{Notice that, altough the $\beta_n$ are generically arbitrary, the choice $\beta_1 + 2 \beta_2+\beta_3=0$ would imply a theory of two \textit{interacting} massless gravitons, thus violating the theorem of \cite{Boulanger:2000rq}. However, for this choice $3$ out of the total $7$ d.o.f. become strongly coupled and the linear expansion around Minkowski breaks then down. CM thanks Xavier Bekaert and Sayed Fawad Hassan for clarifications on this point.} of the latter being given by 
\beq \label{mFP}
m_{\textrm{FP}}^2 \equiv m_g^2 (1+\alpha^2) \,(\beta_1 + 2 \beta_2+\beta_3)\,.
\eeq
It can then be shown that $G_{\mu \nu}$ can be identified with the graviton of general relativity \textit{to all orders} and the full set of \textit{off--shell} mixed interaction vertices involving $G_{\mu \nu}$ and $M_{\mu \nu}$  at cubic and quartic order has been computed, with the first non--trivial interactions appearing at cubic level \cite{Babichev:2016bxi, Babichev:2016hir}.

To compare these off--shell vertices with vertices from string ampitudes, where all external string states are on--shell, in \cite{Lust:2021jps} we imposed that $G_{\mu \nu}$ and $M_{\mu \nu}$ be transverse, traceless, as well as satisfy their equations of motion
\beq
\Box G_{\mu \nu} =0 \quad  , \quad (\Box - m_{\textrm{FP}}^2) M_{\mu \nu} =0
\eeq
and brought the set of all cubic vertices of bimetric theory around Minkowski backgrounds to the form
\begin{subequations}
\begin{align}
 \mathcal{L}_{\textrm{G}^3}^{\textrm{bim}}&= \frac{1}{m_g \sqrt{1+\alpha^2}} \, G^{\mu \nu}\big(\partial_\mu  G_{\rho \sigma} \partial_\nu G^{\rho \sigma}  -2\partial_\nu  G_{\rho \sigma} \partial^\sigma  G_\mu^\rho\big)     \label{bimetric_expansion_1}
 \crbig
 \mathcal{L}_{\textrm{G}^2\textrm{M}}^{\textrm{bim}}&= 0    \label{bimetric_expansion_2}
\crbig
  \mathcal{L}_{\textrm{GM}^2}^{\textrm{bim}}&= \frac{1 }{m_g\sqrt{1+\alpha^2}}\,  \Big[ G^{\mu\nu}\big(\partial_\mu M_{\rho \sigma}  \partial_\nu M^{\rho \sigma}  -4 \partial_\nu M_{\rho \sigma}  \partial^\sigma M_{\mu}^{\rho}   \big) \nonumber
\crbig
& \qquad \qquad + 2 M^{\mu\nu} \big( \partial_\mu G_{\rho \sigma}  \partial_\nu M^{\rho \sigma}-\partial_\rho G_{\mu \sigma}  \partial_\nu M^{\rho \sigma} \big) \Big] \label{bimetric_expansion_3}
\crbig
\mathcal{L}_{\textrm{M}^3}^{\textrm{bim}}&=\frac{ (-\beta_1+\beta_3) \,(1+\alpha^2)^{3/2} m_g}{6\, \alpha}\,   [M^3] \nonumber
\crbig
& \qquad  + \frac{(1-\alpha^2) }{m_g\alpha \sqrt{1+\alpha^2}}  \, M^{\mu \nu}\big(\partial_\mu  M_{\rho \sigma} \partial_\nu M^{\rho \sigma}  -2\partial_\nu  M_{\rho \sigma} \partial^\sigma  M_\mu^\rho\big)    \,,  \label{bimetric_expansion_4}
\end{align}
\end{subequations}
where brackets denote taking the trace over spacetime indices. Notice that the derivative self--interactions of $M_{\mu \nu}$ are identical to those of the GR graviton and that it is only $M_{\mu \nu}$ that enjoys, in addition, non--derivative self--interactions. 

\begin{table}
\centering 
\renewcommand{\arraystretch}{1.5}
  \begin{tabular}{ c || c | c | c  | c }
   $M_{\mu \nu}$ &  $m^2$ & $\frac{1}{M_{\textrm{Pl}}}$ & coupling & coupling    \\ \hline
   bimetric & $m_g^2 (1+\alpha^2) \,(\beta_1 + 2 \beta_2+\beta_3)$ &  $ \frac{1}{m_g \sqrt{1+\alpha^2}} $ & $ \frac{(-\beta_1+\beta_3) \,(1+\alpha^2)^{3/2}}{ 6\alpha} $ & $\frac{1-\alpha^2}{\alpha}$   \\ \hline
   brane $\mathcal{N}=1,2,4$ & $1/\alpha'$ &  ``$g_{\textrm{c}}  $'' & ``$1$'' & ``$2$'' \\ \hline
   bulk $\mathcal{N}=\widetilde 2$, bos & $4/\alpha'$ & $g_{\textrm{c}} $ & $0$ & $0$ \\ \hline
   bulk $\mathcal{N}=\widetilde 4, \widetilde 8$ & $4/\alpha'$ & $g_{\textrm{c}} $ & $0$ & $1$ 
  \end{tabular}
\renewcommand{\arraystretch}{1}
\caption{Summary of results and comparison. The last two columns contain the (dimensionless part of) the couplings associated with non--derivative cubic and $2$--derivative cubic self--interactions of the massive spin--$2$ state respectively. The quotation marks denote that relating the bimetric coefficients with string parameters is impossible in the brane case, due the numerical discrepancies we found in \cite{Lust:2021jps}.} \label{summary}
\end{table}
We are now ready to compare bimetric theory with our results. To begin with, there is no doubt that the graviton self--interactions (\ref{bimetric_expansion_1}) match (\ref{known_2}) for all cases $\mathcal{N}=2,4,8$ (and to lowest order in $\alpha'$ for the bosonic, since $\mathcal{R}^3$--like terms are not present in (\ref{HR})) as expected, upon imposing
\beq \label{coupling}
 \frac{1}{m_g \sqrt{1+\alpha^2}} \overset{!}{=} g_{\textrm{c}}\,.
\eeq
 Next, we identify $M_{\mu \nu}$ with the massive spin--2 states (\ref{dc_massive_1})--(\ref{dc_massive_4}) we have constructed, noting that, since in (\ref{HR}) none of the spin--$2$ tensors are charged under non--abelian symmetries, we also restrict ourselves to non--BPS states, namely cases $\mathcal{N}=\widetilde 2, \widetilde 4, \widetilde 8$, as explained in the previous section. The identification then imposes that their masses be equal to (\ref{mFP}), namely
\beq
m_g^2 (1+\alpha^2) \,(\beta_1 + 2 \beta_2+\beta_3) \overset{!}{=} \frac{4}{\alpha'}\,.
\eeq
(\ref{bimetric_expansion_2}) then matches (\ref{new_1b}) in all cases: there can be no vertices involving a single massive spin--$2$ state and two gravitons in all cases. Interestingly, this observation has been argued to be a \textit{discriminatory} property of ghost--free bimetric theory \cite{Babichev:2016bxi}. However, the generic field theory results of \cite{Arkani-Hamed:2017jhn} and our string results serve as evidence that this a universal property of field and perhaps also string theories. Next, we also observe that (\ref{bimetric_expansion_3}) \textit{matches} (\ref{new_2b}) for all cases $\mathcal{N}=\widetilde 2, \widetilde 4, \widetilde 8$ (and to lowest order for the bosonic) under the condition (\ref{coupling}). Crucially, this was not the case for the brane massive spin--2 we used in \cite{Lust:2021jps}.

Next, comparison of (\ref{bimetric_expansion_4}) with (\ref{one_van_ampl}) for the $\mathcal{N}=\widetilde{2}$ case and with (\ref{ampl_2_bosonic_4}) for the identical bosonic case implies (since strictly $\alpha>0$)
\beq
\beta_1\overset{!}{=} \beta_3 \quad and \quad \alpha \overset{!}{=} 1
\eeq
to guarantee the absence of all cubic vertices involving three massive spin--2 states. However, recalling from (\ref{HR}) that $\alpha$ measures the relative strength of the gravitational interactions mediated by the two metrics $g_{\mu \nu}$ and $f_{\mu \nu}$ and inspecting the fluctuations (\ref{eigen}), we deduce that a value of $\alpha$ close to $1$ implies that $G_{\mu \nu}$ can no longer be uniquely identified with the graviton, so it cannot be uniquely associated with the metric of spacetime in which $M_{\mu \nu}$ propagates. It is further worth noting that the limit of ghost--free bimetric theory in which general relativity is restored is achieved for $\alpha \ll 1$ \cite{Baccetti:2012bk, Hassan:2014vja, Akrami:2015qga}. We conclude that the massive spin--2 states of the $\mathcal{N}=2$ and of the bosonic case we have constructed \textit{cannot} be interpreted as the massive spin--2 mode of bimetric theory at cubic level. 

To continue, we compare with the $\mathcal{N}=\widetilde 4, \widetilde 8$ cases, namely (\ref{bimetric_expansion_4}) with (\ref{massive_3_fin}). We find a match subject to the conditions (\ref{coupling}) and 
\beq \label{param_cons}
\beta_1= \beta_3 \quad , \quad \alpha \approx 0.62 \,,
\eeq
where the first is a consequence of the absence of non--derivative self--interactions of the massive spin--2 in (\ref{massive_3_fin}) and the second is due to solving the equation $\alpha^2 +\alpha-1=0$ (strictly for $\alpha>0$), that is enforced by the derivative self--interactions. Since this value of $\alpha$ is smaller than $1$, we accept it as a solution and conclude that non--BPS states originating in the massive spin--2 states of $\mathcal{N}=4, 8$ we have constructed \textit{can} be interpreted as the massive spin--2 mode of bimetric theory at cubic level; we remind the reader that the original states are charged under $R$--symmetry, which is why we cannot preserve the full $\mathcal{N}=4,8$ to achieve this match. Moreover, we notice that, despite the match, our non--BPS states cannot serve as viable dark matter candidates, since for $M_{\mu \nu}$ of bimetric theory to be such, $\alpha$ has systematically been constrained to the range \cite{Babichev:2016bxi, Babichev:2016hir}
\beq
10^{-15} \lesssim \alpha  \lesssim10^{-11}.
\eeq
In table \ref{summary} we summarise our results and compare them with our previous work \cite{Lust:2021jps}.

Finally, let us conclude with a remark on the first of conditions (\ref{param_cons}). Around Minkowski backgrounds as we have assumed, ghost--free bimetric theory is a $3$--parameter family of theories, since the vanishing of the cosmological constant of the two metrics forces \cite{Hassan:2011zd}
\beq
\beta_0+3\beta_1+3\beta_2+\beta_3 = \beta_1+3 \beta_2+3 \beta_3+\beta_4 =0\,,
\eeq
namely $2$ $\beta_n$'s can be fixed. Moreover, canonically normalising the Fierz--Pauli mass (\ref{mFP}) further forces
\beq
\beta_1 + 2 \beta_2+\beta_3=1\,,
\eeq
which reduces the number of free (dimensionless) parameters to $2$ \cite{Bonifacio:2017nnt}. Due to the first of conditions (\ref{param_cons}), our cubic string amplitudes then further select
\beq \label{cons2}
\beta_0 = \beta_4\,,
\eeq
namely a $1$--parameter subfamily of bigravities. Interestingly, these conditions on the $\beta_n$, namely the first of (\ref{param_cons}) and (\ref{cons2}), imply that the action (\ref{HR}) enjoys a $\mathbb{Z}_2$ symmetry under interchange of $g_{\mu \nu}$ and $f_{\mu \nu}$ along with their respective Planck masses and also correspond to a special $1$--parameter subfamily of the $2$--parameter family of massive dRGT gravities singled out by imposing the absence of asymptotic superluminality \cite{Bonifacio:2017nnt}. Intriguingly, the very same subfamily is selected by the cubic order of the double copy of dRGT massive gravity in terms of Yang--Mills fields with Proca--like terms \cite{Momeni:2020vvr, Johnson:2020pny} and also, independently, by the scenario in which the massive spin--$2$ of dRGT to cubic order is the highest--spin component of an $\mathcal{N}=4$ supermultiplet \cite{Engelbrecht:2022aao}.

\section{Concluding remarks}

In this work, we have derived a double copy construction of massive gravitons from and within string theory. Specifically, we used the product of two massive vectors of the open string spectrum, which appear in compactifications to four dimensions, to construct massive spin--$2$ tensors as closed string states. This is in analogy to the structure of the massless closed string graviton as the product of two massless open string vector states. As we have explained, the massive open string vectors and the massive closed string spin--2 tensors do not belong to the leading Regge trajectory, but are the lowest--lying massive string excitations of the next subleading Regge trajectory. As we have shown, the $3$--point string amplitudes of three such massive open string vectors or of the corresponding closed string spin--2 tensors are only non-vanishing for the case of $\mathcal{N}=2, 4$ supersymmetry in the open string sector, namely \textit{extended} brane supersymmetry, respectively $\mathcal{N}= 4,8$ bulk supersymmetry in the closed string double copy construction. For the generic case of open string $\mathcal{N}= 1$ supersymmetry, respectively for $\mathcal{N}= 2$ supersymmetry in the closed string bulk, these amplitudes are zero.

We note that it is further possible to construct massive vectors of the leading open string Regge trajectory by replacing in (\ref{massive_graviton_open}) one spacetime index $\mu$ or $\nu$ with an internal index. However, one can show that the zero modes of these massive open string vectors have vanishing  $3$--point couplings. On the other hand, relaxing the assumption (\ref{no_leak}), the Kaluza--Klein excitations of these states, i.e. the massive vectors with additional internal momenta, possess non-vanishing $3$--point couplings; these can simply be obtained by replacing one spacetime momentum by an internal momentum in the $3$--point couplings \cite{Lust:2021jps}  of the corresponding spin--$2$ massive open string states, see also the comment on breaking the $R$--symmetry below (\ref{charge_4}).

With the computation of $3$--point amplitudes of one massive spin--$2$ bulk state and two gravitons, of two massive spin--$2$ bulk states and one graviton and of three massive bulk spin--$2$ bulk states yielding results \textit{finite} in $\alpha'$, we were able to \textit{reproduce the cubic} Lagrangian of ghost--free bimetric theory around flat spacetime, for bulk states originating in multiplets of $\mathcal{N}= 4,8$ supersymmetry. To facilitate the comparison, given that the metric tensors of bimetric theory carry no nonabelian charges, while our bulk states transform a priori nontrivially under the respective $R$--symmetry by construction, we chose by hand the corresponding singlet representations, thereby not preserving the full bulk supersymmetry and rendering the bulk states non--BPS. Moreover, the agreement is achieved by matching the string parameters to those of the effective bimetric theory. In particular, we relate $\alpha'$, or equivalently the closed string coupling, to the Planck masses of the two metrics of bimetric theory and determine the ratio of the latter; we further constrain the dimensionless parameters of the bimetric potential to a $1$--parameter family of theories. 

Intriguingly, the latter matches the one singled out by the absence of asymptotic superluminality in massive gravity \cite{Bonifacio:2017nnt}, by the cubic level of the field theory double copy of dRGT massive gravity \cite{Momeni:2020vvr, Johnson:2020pny} and also by supersymmetrisations of dRGT, with the massive spin--$2$ assumed to be the highest--spin component of an $\mathcal{N}=4$ supermultiplet \cite{Engelbrecht:2022aao}, unlike our own bulk states, that originate in multiplets that further accommodate higher spins. Moreover, our match with bimetric theory is in contrast  to our previous work \cite{Lust:2021jps}, where we showed that the $3$--point amplitudes of the lowest--lying brane massive string states, that belong to the leading Regge trajectory, cannot be matched to the couplings of the bimetric theory. It should be highlighted that, since our string states and corresponding amplitudes are on--shell, all our statements refer to a match with the on--shell part of the cubic bigravity action. Higher--point amplitudes deserve future investigation, as they have the potential to shed light on the off--shell structure of spin--2 string interactions and also on the subtleties of the realisation of our stringy double copy construction at higher orders, where, at least in the field theory side in four dimensions, the massive double copy may be plagued by spurious poles \cite{Momeni:2020vvr, Johnson:2020pny}. To conclude, at least at the level of on--shell cubic interactions, among the lowest--lying spin--$2$ string states in four dimensions, it is only bulk and not brane states that may enjoy self--interactions that are identical to those of the graviton and, in addition, only the closed string double copy construction of massive gravity in string theory leads to bimetric theory as its ``effective'' description.

\paragraph{Acknowledgments:}
C.M. thanks Carlo Angelantonj, Constantin Bachas, Nicolas Boulanger and Anastasios Petkou for useful discussions, Renann Lipinski Jusinskas and Oliver Schlotterer for pivotal insights and discussions and is especially grateful to Evgeny Skvortsov for several enlightening discussions of formative impact. She further thanks CEICO of the Institute of Physics of the Czech Academy of Sciences and the organisers of the 12th Crete Regional Meeting in String Theory and respectively of ``QCD meets Gravity 2022'', where she presented part of this work. The work of D.L.~is supported by the Origins Excellence Cluster and by the German-Israel-Project (DIP) on Holography and the Swampland. The work of C.M. is supported by the European Research Council (ERC) under the European Union’s Horizon 2020 research and innovation programme (grant agreement No 101002551) and was partially supported by the Fonds de la Recherche Scientifique -- FNRS under Grants No. F.4544.21 (HigherSpinGraWave) and No. R.M005.19.

\appendix 
\section{Cardy's trick and the string spectrum}\label{conv_app}

Following for example \cite{Polchinski:1998rr}, in a 2D CFT of a bosonic field $X(z,\ov{z})$, the two--point function on the sphere $S_2$ (or, equivalently, the infinite complex plane) takes the form
\beq
\langle X(z,\ov{z})X(w,\ov{w}) \rangle_{S^2} = - \frac{\alpha'}{2} \ln |z-w|^2\,,
\eeq
while on the disk $D_2$ (or equivalently on the UHP $\mathcal{H}_+$) it receives an image contribution because of the presence of the boundary:
\beq
\langle X(z,\ov{z})X(w,\ov{w}) \rangle_{D_2}  = - \frac{\alpha'}{2} \big\{\ln |z-w|^2+\ln |z-\ov{w}|^2 \big\} \,,
\eeq
where Neumann boundary conditions are assumed at both ends of the open string. Since $S_2$ and $D_2$ are the relevant worldsheet topologies for closed and open strings respectively, for amplitudes involving solely closed string states as external states, one has that
\beq \label{OPE_closed}
\langle X^m(z) X^n(w) \rangle = -\frac{\alpha'}{2} \eta^{m n} \, \ln(z-w) \,,
\eeq
where  $X(z,\ov{z})=X(z)+\widetilde{X}(\ov{z}) $ on--shell, while if the external states are solely open string states, whose vertex operators are thought of as being inserted on the real line $\mathbb{R}$, the OPE simplifies to
\beq \label{OPE_open}
\langle X^m(x) X^n(y) \rangle = - 2 \alpha' \eta^{m n} \, \ln(x-y) \quad , \quad x,y \in \mathbb{R}\,.
\eeq
A comparison of the form of (\ref{OPE_closed}) and of (\ref{OPE_open}) has the following implications:
\begin{enumerate}
\item it suggests the convenient choices \cite{Kawai:1985xq}
\beq
\begin{array}{ccl}
\alpha'=\begin{cases}
  2\,, &  \text{ for closed strings} \\
  1/2\,, &  \text{ for open strings}\,,
\end{cases}
\end{array}
\eeq
which we nevertheless refrain from in this work, since $\alpha'$ must be kept arbitrary for our purposes as emphasised in the main text.
\item (\ref{OPE_closed}) and (\ref{OPE_open}) are equivalent upon the replacement
\beq \label{doubling}
X(x) \rightarrow 2X(z) \quad , \quad x \rightarrow z\,.
\eeq
Moreover, in the CFT ``bulk'', the energy--momentum tensor takes the form
\beq \label{T_closed}
 T_{X}(z)= -\frac{1}{\alpha'} \, \partial X \partial X\,(z)\,,
\eeq
which on the boundary simplifies to
\beq \label{T_open}
 T_{X}(x)= -\frac{1}{4\alpha'} \, \partial X \partial X\,(x)
\eeq
due to the inverse of (\ref{doubling}). Since closed and open string states are thought of as being inserted in the ``bulk'' and at the boundary respectively, the suitable momentum eigenstates are correspondingly $e^{ikX(z,\ov{z})}$ and $e^{ikX(x)}$. The conformal dimension of the latter can then be computed in two equivalent ways: either one uses (\ref{T_open}) and (\ref{OPE_open}), or one performs the substitution (\ref{doubling}) in the exponential and uses (\ref{T_closed}) and (\ref{OPE_closed}), since the full disk of the boundary CFT is available to open string states.
\end{enumerate}

\begin{table}
\centering 
\renewcommand{\arraystretch}{1.5}
  \begin{tabular}{ c || c | c | c   }
   type & momentum eigenstate &  eigenstate's conf. dimension & $M_N^2$   \\ \hline
   closed & $e^{ikX(z,\ov{z})}=e^{ikX(z)+ik\widetilde{X}(\ov{z})}$ & $\big( \alpha' k^2/4, \alpha' k^2/4 \big)$  & $4N/\alpha'$   \\ \hline
   open  & $e^{ikX(x)}=e^{2ikX(z)}$  & $\alpha' k^2$ & $N/\alpha'$ 
  \end{tabular}
\renewcommand{\arraystretch}{1}
\caption{Closed vs open string spectra} \label{closed_open}
\end{table}
As documented in table \ref{closed_open}, it is then easy to see that the conformal dimension of open string eigenstates is four times the conformal dimension of the holomorphic part of closed string eigenstates with the same momentum. This further means that:
\begin{itemize}
\item the mass spacing of the closed string spectrum is \textit{four} times the one of the open string spectrum, and
\item the momentum eigenstate of a closed string state with momentum $k^m$ can be written as
\beq
e^{ikX(z,\ov{z})} = e^{ikX(z)}e^{ik\widetilde{X}(\ov{z})} \quad \overset{(\ref{doubling})}{\longleftarrow} \quad \big( e^{i\frac{k}{2}X(x)} \big)_{\textrm{L}}  \, \big( e^{i\frac{k}{2}X(y)} \big)_{\textrm{R}}\,,
\eeq
namely as product of two open string momentum eigenstates, each of which can be thought of as corresponding to a string carrying momentum $p^m=\frac{1}{2}k^m$. This is necessary in order to correctly reproduce the conformal dimensions of both closed and open string eigenstates in the ``bulk'' and in the boundary CFT respectively; it should be obvious that this consideration is important for massive states yet redundant for massless ones. Another, well--known, way of seeing this is via the Fourier expansion of $X(z,\ov{z})$, which implies that its momentum is double that of the left (or right) movers $X(z)$ (or $\widetilde{X}(\ov{z})$). 
\end{itemize}

This essentially means that the map from open to closed strings can be thought of as
\beq
\alpha' \rightarrow \frac{1}{4} \alpha'\,.
\eeq
The above considerations essentially reflect Cardy's trick \cite{Cardy:1989ir} and are valid for both the bosonic string and the superstring, not affecting fermions and the ghost systems. Since we do not consider interactions of open and closed strings, for either kind we use $z \in \mathbb{C}$ as the worldsheet coordinate throughout this work. 

\section{OPEs} \label{app_ope}

Following for example \cite{Blumenhagen:2013fgp}, we have that for open strings
\beq
\begin{array}{ccl}
 i\partial X^m (w)  e^{ipX(z)}& =& \Big[ \frac{2\alpha' p^m}{w-z}+i\partial X^m(z)+i(w-z)\partial^2X^m(z)+\dots \Big]\,e^{ipX(z)}
 \crbig
  T(w) \, e^{ipX(z)} &=& \Big[\frac{\alpha'p^2}{(w-z)^2} +\frac{\partial_z}{w-z}-\frac{1}{4\alpha'}\partial X(z) \partial X(z) 
  \crbig
&& \quad -\frac{1}{4\alpha'}(w-z) \partial X(z) \partial^2 X(z) + \dots   \Big]\,e^{ipX(z)}
\end{array}
\eeq
while for closed strings
\beq
\begin{array}{ccl}
  i\partial X^m (w)  e^{ikX(z,\ov{z})}& =& \Big[ \frac{1}{2}\frac{\alpha' k^m}{w-z}+i\partial X^m(z)+i(w-z)\partial^2X^m(z)+\dots \Big]\,e^{ikX(z,\ov{z})}
\crbig
 T(w) \, e^{ikX(z,\ov{z})} &=& \Big[\frac{1}{4}\frac{\alpha'k^2}{(w-z)^2} +\frac{\partial_z}{w-z}-\frac{1}{\alpha'}\partial X(z) \partial X(z) 
  \crbig
&& \quad -\frac{1}{\alpha'}(w-z) \partial X(z) \partial^2 X(z) + \dots   \Big]\,e^{ikX(z,\ov{z})} \,.
\crbig
\end{array}
\eeq
In both cases,
\beq \label{OPE_exp}
\begin{array}{ccl}
 \psi^\mu(z) \psi^\nu(w)  & =& \frac{\eta^{\mu \nu}}{z-w} + \psi^\mu \psi^\nu (w) +(z-w) \, \partial \psi^\mu \psi^\nu (w) + \dots
 \crbig
  e^{\phi(z)} e^{-\phi(w)} &=& (z-w)+(z-w)^2 \,\partial \phi(w)+\dots
 \crbig
 \eta(z) \xi(w) &= & \frac{1}{z-w}+\eta(w)\xi(w)+(z-w) \partial \eta(w) \xi(w)+\dots
\end{array}
\eeq
All relations given in the appendices are equally well valid in the critical and in the compactifications to four dimensions that we consider in this work.

\end{document}